\documentclass[preprint,prc,aps,tightenlines,floats,showpacs,showkeys,superscriptaddress,nofootinbib]{revtex4-1}
\usepackage[dvips]{graphicx,color}
\usepackage{dcolumn}
\usepackage{epsfig}
\usepackage{bm}
\usepackage[dvipdfm,colorlinks=true,linkcolor=blue,filecolor=blue,urlcolor=blue]{hyperref}

\newcommand{\ket}[1]{|{#1}\rangle}
\newcommand{\bra}[1]{\langle{#1}|}

\newcommand{\slas}[1]{\not\!{#1}}
\newcommand{\sla}[1]{\not\!\!{#1}}

\begin{document}
\title{$\Lambda (1405)$ photoproduction based on chiral unitary model}
\author{S.X. Nakamura}
\thanks{Present address: 
Department of Physics, Osaka University, Toyonaka, Osaka 560-0043, Japan}
\email{nakamura@kern.phys.sci.osaka-u.ac.jp}
\affiliation{Yukawa Institute for Theoretical Physics, Kyoto University, Kyoto, 606-8502, Japan}
\author{D. Jido}
\affiliation{Department of Physics,
Tokyo Metropolitan University,
Hachioji, Tokyo 192-0397, Japan}

\begin{abstract}
Recent CLAS data for the $\pi\Sigma$ invariant mass distributions (line-shapes)
in the $\gamma p \to K^{+} \pi \Sigma$ reaction are theoretically investigated. 
The line-shapes have peaks associated with the $\Lambda(1405)$ excitation.
Our model consists of gauge invariant photo-production mechanisms, and
the chiral unitary model that gives
the rescattering amplitudes where $\Lambda(1405)$ is contained.
It is found that, while the $\pi\Sigma$ line-shape data in the
$\Lambda(1405)$ region are successfully reproduced by our model for all the charge
states, the production mechanism is not so simple that we need to
introduce parameters associated with short-range dynamics to fit the data.
Our detailed analysis suggests that the nonresonant background
contribution is not negligible, and its sizable effect shifts
the $\Lambda(1405)$ peak position by several MeV.
We also analyze the data using a Breit-Wigner amplitudes instead of
those from the chiral unitary model.
We find that the fitted Breit-Wigner parameters
are closer to the higher pole position for 
$\Lambda(1405)$ of the chiral unitary model. 
This work sets a starting point for a fuller analysis in which 
line-shape as well as $K^+$ angular distribution data
are simultaneously analyzed for
extracting $\Lambda(1405)$ pole(s).
\end{abstract}

\maketitle

\section{introduction}
\label{sec:intro}

The pole structure of the $\Lambda(1405)$ resonance is a key issue to understand 
the nature of $\Lambda(1405)$ and the $\bar KN$ interaction. 
Because $\Lambda(1405)$ decays exclusively into the $\pi \Sigma$ channel with $I=0$  
by the strong interaction, 
a signal associated with $\Lambda(1405)$ is expected to be observed in the $\pi\Sigma$ 
invariant mass distributions (to be referred to as ``line-shape'') of
certain $\Lambda(1405)$ production reactions.  

In old bubble chamber experiments, 
bumps associated with the $\Lambda(1405)$ excitation have been 
observed in the $\pi\Sigma$ line-shapes of hadron induced reactions, 
such as $\pi^{-} p \to K^{+} \pi \Sigma$ \cite{Thomas:1973uh},
$K^{-}p \to \pi^{-} \pi^{+} \pi \Sigma$~\cite{Hemingway:1984pz} and
$K^{-}d \to n \pi^{+} \Sigma^{-}$ \cite{Braun:1977wd}. 
The observed bumps in the first two 
experiments are consistent with the $\Lambda(1405)$ resonance at 1405 MeV, while 
the reaction with a deuteron target found the $\Lambda(1405)$ resonance 
at 1420 MeV. 
Recently, 
the $\pi^{0}\Sigma^{0}$ line-shape for the $\Lambda(1405)$ energies 
was also measured in hadronic reactions, such as $K^{-}p \to \pi^{0} \pi^{0} \Sigma^{0}$
with 514-750 MeV/c kaon momenta
by Crystal Ball Collaboration~\cite{Prakhov:2004an},
and $pp \to p K^{+} \Lambda(1405)$ with 3.65 GeV/c proton beam 
at COSY-J\"ulich~\cite{Zychor:2007gf}. 

Although there have been several data for the $\Lambda(1405)$ spectrum
from the hadron beam experiments as mentioned above, 
the quality of the data is still not sufficient for extracting $\Lambda(1405)$ pole(s).
The situation has been changed by recent photon-beam experiments.
The first photoproduction of the $\Lambda(1405)$ resonance was observed 
at SPring 8 by LEPS collaboration in the $\gamma p \to K^{+} \pi \Sigma$
reaction with the photon 
energy of 1.5-2.4 GeV~\cite{Ahn:2003mv,Niiyama:2008rt}. In this experiment, 
the $\pi^{-}\Sigma^{+}$ and $\pi^{+} \Sigma^{-}$ line-shapes
were measured, and they were found to be different from
each other, owing to the 
interference between the $I=0$ resonant and $I=1$ non-resonant contributions. 
A high statistics, wide angle coverage experiment for
the $\gamma p \to K^{+} \pi \Sigma$ reaction 
was performed at Jefferson Laboratory by the CLAS collaboration,
for center-of-mass energies $1.95 < W < 2.85$ GeV~\cite{Moriya:2013eb,Moriya:2013hwg}.
In this experiment, all the three charge states of the $\pi \Sigma$
channels were simultaneously observed in the $\gamma p$ scattering
for the first time, 
and the differential cross sections were measured 
for the $\pi\Sigma$ line-shape
and for the $K^+$ angular distribution.
This is the cleanest data that cover the kinematics of $\Lambda(1405)$
excitation, which encourages theorists to seriously work on extracting
the $\Lambda(1405)$ pole(s) from data for the first time.
Very recently the spectral shape of $\Lambda(1405)$ has been also observed
in electroproduction in the range of $1.0 < Q^{2} < 3.0$ $({\rm GeV}/c)^{2}$~\cite{Lu:2013nza}.

The coupled-channel approach based on 
the chiral effective theory (chiral unitary model) suggests that the $\Lambda(1405)$ resonance
is composed of two poles located between the $\bar KN$ and $\pi \Sigma$ 
thresholds~\cite{Hyodo:2011ur} and these states have different masses, widths
and couplings to the $\bar KN$ and $\pi\Sigma$ channels. One pole 
is located at $1426- 16i$ MeV with a dominant coupling to $\bar KN$, while
the other is sitting at $1390-66i$ MeV with a strong coupling to $\pi \Sigma$~\cite{Jido:2003cb}.
These two states are generated dynamically by the attractive interaction in 
the $\bar KN$ and $\pi\Sigma$ channels with $I=0$~\cite{Hyodo:2007jq}.
Because the $\Lambda(1405)$ resonance is composed of two states 
which have different weight to couple with $\bar KN$ and $\pi \Sigma$,
the spectral shape of the $\pi\Sigma$ line-shape in the $\Lambda(1405)$ region
depends on how $\Lambda(1405)$ is produced, as pointed out in Ref.~\cite{Jido:2003cb}.  
Reference~\cite{Jido:2003cb} predicts that the $\Lambda(1405)$
resonance in the $\bar KN \to \pi \Sigma$ channel has a peak at 1420 MeV with a 
narrower width because the higher pole strongly couples to the $\bar KN$ channel. 
The study of Ref.~\cite{Jido:2009jf} showed that, in the $K^{-}d \to n \pi \Sigma$
reaction, $\Lambda(1405)$ is dominantly produced by $\bar KN$,
and the $\pi\Sigma$ line-shape
has a peak at 1420 MeV as seen in the old
bubble chamber experiment~\cite{Braun:1977wd}. 

It is important to confirm the two-pole structure by analyzing the new
CLAS data, and if so,
it will be interesting to see how the two-pole structure plays a role in
the $\pi\Sigma$ line-shape.
In order to extract the $\Lambda(1405)$ resonance pole(s) from
the production data, we develop a model that consists of production
mechanism followed by the final state interaction (FSI);
$\Lambda(1405)$ is excited in the FSI.
Through a careful analysis of the data, we can pin down 
the production mechanism as well as the scattering amplitude
responsible for the FSI.
Then the $\Lambda(1405)$ pole(s) will be extracted from the scattering amplitude.
Such an analysis of the new CLAS data has been done in
Refs.~\cite{roca1,roca2} using a simple production mechanism.

In this work, we focus on the photoproduction of $\Lambda(1405)$
in $\gamma p \to K^{+} \pi \Sigma$, and
investigate the new CLAS data for the $\pi\Sigma$ line-shape~\cite{Moriya:2013eb}.
The first study of the reaction was done in Ref.~\cite{nacher}, in which 
a simple diagram was considered for the $\Lambda(1405)$ production mechanism 
and the $\Lambda(1405)$ is described by the chiral unitary approach. 
Related calculations were also done in Refs.~\cite{nacher2,borasoy}.
Although the calculation of Ref.~\cite{nacher} was to get a rough
estimate of the cross sections, in the advent of the fairly precise
data, it is necessary to develop a quantitative model to extract 
the $\Lambda(1405)$ properties from the data.
In this work, we extend and refine the model of Ref.~\cite{nacher} by
considering more production mechanisms that are gauge invariant at the
tree-level.
We consider relevant meson-exchange mechanisms, and contact terms that
simulate short-range mechanisms. 
We explain details of the model, and successfully fit the CLAS data with it. 
Then we discuss a role played by each mechanism, effects of non-resonant
contributions, and a possibility of a single-pole solution of $\Lambda(1405)$.
By doing so, we set a starting point for a full analysis in which we
simultaneously analyze the data for line-shape~\cite{Moriya:2013eb}
and the $K^+$ angular distribution~\cite{Moriya:2013hwg} to study the
pole structure of $\Lambda(1405)$.
Such a full analysis is left to a future work.
We expect the $K^+$ angular distribution data are an important
information to pin down the production mechanism.

The rest of this paper is organized as follows:
We give a detailed description of our model in Sec.~\ref{sec:model}.
Then we show numerical results and discuss them in
Sec.~\ref{sec:result}, followed by a summary in Sec.~\ref{sec:summary}.
Expressions for Lagrangians and photo-production operators,
and also model parameters are collected in Appendices.

\section{Model}
\label{sec:model}

\subsection{Kinematics and cross section formula}
\label{sec:xs}

First we define kinematical variables.
We consider the $\gamma (q) + p (p) \to K^+(k) + \pi (k_\pi) + \Sigma (p')$
reaction in which the variables in the parentheses are four-momenta for
the particles in the total center-of-mass system.
The differential cross section for the reaction is derived following a
standard procedure, and given as
\begin{eqnarray}
\label{eq:xs}
d\sigma = 
{M_p M_\Sigma
\lambda^{1/2}(s,M^2_{\pi\Sigma},m^2_{K^+})
\lambda^{1/2}(M^2_{\pi\Sigma},M^2_\Sigma,m^2_\pi) 
\over 
 512 \pi^5 s (s-M_p^2) M_{\pi\Sigma}} 
\sum_{\rm spin}
|T_{K^+\pi\Sigma,\gamma p}|^2 
d M_{\pi\Sigma} d\Omega_k d\Omega^*_{p'}
\ ,
\end{eqnarray}
where $M_p$, $M_\Sigma$, $m_{K^+}$ and $m_\pi$ are the masses of the
proton, $\Sigma$, $K^+$ and $\pi$, respectively, and the K\"allen
function is denoted by $\lambda(x,y,z)$.
The symbol $s$ is
the squared total energy of the system, and is related to the four-momenta by 
$s = W^2 = (q+p)^2 = (k+k_\pi+p')^2$,
while the invariant mass of the $\pi\Sigma$ subsystem is
$M^2_{\pi\Sigma} = (k_\pi+p')^2$.
The kinematical variables with asterisk 
stand for the quantities in
the $\pi\Sigma$ center-of-mass system.
The summation of spin and polarization states in initial and final particles are
indicated by $\sum_{\rm spin}$; the average factor, 1/4, for the initial
states is already included in the factor of the formula.  
All information about the dynamics is encoded into the reaction
amplitude $T_{K^+\pi\Sigma,\gamma p}$ in Eq.~(\ref{eq:xs}), and is discussed in detail in the next
subsection.
The line-shape of the $\pi\Sigma$ spectrum is obtained by integrating Eq.~(\ref{eq:xs})
over the angular part of $\vec p'^*$ and $\vec k$, and given as
\begin{eqnarray}
\label{eq:ls}
{d\sigma \over d M_{\pi\Sigma}} 
= 
\sum_{\rm spin}
\int  d\Omega_k d\Omega^*_{p'}
{M_p M_\Sigma
\lambda^{1/2}(s,M^2_{\pi\Sigma},m^2_{K^+})
\lambda^{1/2}(M^2_{\pi\Sigma},M^2_\Sigma,m^2_\pi) 
\over 512 \pi^5 s (s-M_p^2)
 M_{\pi\Sigma}} 
|T_{K^+\pi\Sigma,\gamma p}|^2 
\ .
\end{eqnarray}

\subsection{Photo-production mechanism}
\label{sec:ppm}

As stated in the introduction, 
we describe the $\gamma p \to K^+\pi\Sigma$ reaction by a set of
tree-level mechanisms for $\gamma p \to K^+M_jB_j$ 
($M_jB_j$ : a set of meson and baryon)
followed by $M_jB_j\to \pi\Sigma$ rescattering.
We use an index $j=1,...,10\,$ to specify 
$M_jB_j=K^-p,\bar K^0 n, \pi^0\Lambda, \pi^0\Sigma^0, \eta\Lambda, \eta\Sigma^0$,
$\pi^+\Sigma^-, \pi^-\Sigma^+, K^+\Xi^-, K^0\Xi^0$,  respectively.
Thus the reaction amplitude introduced in Eq.~(\ref{eq:xs}), 
$T^j\equiv T_{K^+M_jB_j,\gamma p}$, is given by 
\begin{eqnarray}
\label{eq:tamp}
T^j = \sum_\alpha V^j_\alpha + T^j_{R}\ ,
\end{eqnarray}
where $V^j_\alpha$ is a tree-level photo-production mechanism.
In the next paragraph, we specify the tree mechanisms that go into our calculation.
The summation of $\alpha$ runs over all of the tree-level photoproduction mechanisms included in our calculation.
Contribution from the rescattering is denoted by $T^j_{R}$.
The rescattering amplitude is calculated with a partial wave expansion
with respect to the relative motion of $M_jB_j$, and 
$(J,L)=(1/2,0)$ and $(1/2,1)$ partial waves are considered; $J$ and $L$
are the total and orbital angular momenta for $M_jB_j$.
The partial wave amplitude is given, with the on-shell factorization, by
\begin{eqnarray}
\label{eq:pw}
T^j_{R;JL} =
\sum_{\alpha}\,\sum_{j'}\, T^{jj'}_{JL}(M^2_{\pi\Sigma})\, G^{j'}_\alpha (M^2_{\pi\Sigma})\, V^{j'}_{\alpha;JL}
\ ,
\end{eqnarray}
where 
$T^j_{R;JL}$ and $V^{j}_{\alpha;JL}$
are partial wave amplitudes of
$T^j_{R}$ and $V^j_\alpha$, respectively, 
and are calculated with the on-shell momenta of relevant particles.
More details about the partial wave expansion, including the relation
between $T^j_{R;JL}$ and $T^j_{R}$, are given in Appendix~\ref{app:pw}.
For the $M_{j'}B_{j'}\to M_jB_j$ scattering amplitudes $T^{jj'}_{JL}$,
we use those from the chiral unitary model
given in Ref.~\cite{ORB} for $(J,L)=(1/2,0)$ wave, and 
in Ref.~\cite{JOR} for $(J,L)=(1/2,1)$ wave.
The $(J,L)=(1/2,0)$ wave contains $\Lambda(1405)$ as double poles, 
while the $(J,L)=(1/2,1)$ wave does not include any resonance and
provide a smooth background.
It is turned out that the contribution from 
the $(J,L)=(1/2,1)$ wave rescattering is small.
We use the meson-baryon Green function, $G^j_\alpha$, calculated with the
dimensional regularization as follows:
\begin{eqnarray} 
G^j_\alpha (s)
 &=& \frac{2 M_{B_j}}{16 \pi^2} \left\{ a^j_\alpha(\mu) + \ln
\frac{M_{B_j}^2}{\mu^2} + \frac{M_{M_j}^2-M_{B_j}^2 + s}{2s} \ln \frac{M_{M_j}^2}{M_{B_j}^2} 
\right. \nonumber \\ & &  
+ \frac{\bar{q}_{j}}{\sqrt{s}} 
\left[ 
\ln(s-(M_{B_j}^2-M_{M_j}^2)+2\bar{q}_{j}\sqrt{s})+
\ln(s+(M_{B_j}^2-M_{M_j}^2)+2\bar{q}_{j}\sqrt{s}) \right. \nonumber  \\
& & \left. 
 - \ln(-s+(M_{B_j}^2-M_{M_j}^2)+2\bar{q}_{j}\sqrt{s})-
\ln(-s-(M_{B_j}^2-M_{M_j}^2)+2\bar{q}_{j}\sqrt{s}) \right]
\Bigg\} \ ,
\label{eq:gpropdr}
\end{eqnarray}        
where $M_{B_j}$ and $M_{M_j}$ are the masses of a baryon $B_j$ and a meson ${M_j}$,
respectively, and we use the values listed in the Particle Data
Group~\cite{pdg} for the masses.
The relative on-shell momentum of $M_jB_j$ corresponding to $s$ is denoted
by $\bar{q}_{j}$.
The symbol $a^j_\alpha(\mu)$ is
the subtraction constant for the regularization scale $\mu$,
and we set $\mu=630$~MeV for all channels.
The subtraction constants can depend on a channel $j$ as well as a
production mechanism $\alpha$;
we will come back to this point at the end of this section.

\begin{figure}
\includegraphics[clip,width=0.80\textwidth]{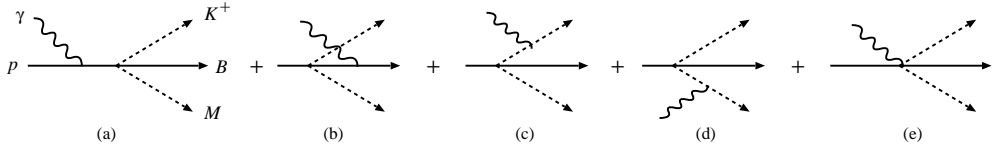}
\caption{\label{fig:wt}
The gauged Weinberg-Tomozawa terms.
}
\end{figure}

\begin{figure}
\includegraphics[clip,width=1.0\textwidth]{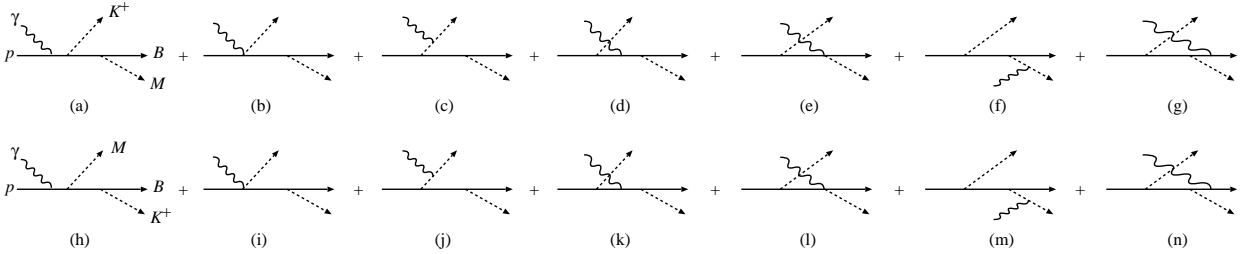}
\caption{\label{fig:br1}
The gauged Born terms.
From the upper row to lower row, $K^+$ and $M$ are exchanged.
}
\end{figure}

\begin{figure}
\includegraphics[clip,width=0.75\textwidth]{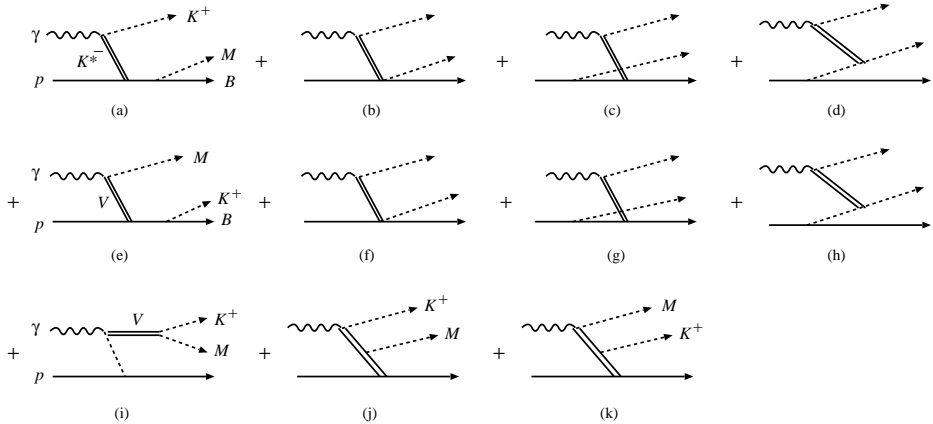}
\caption{\label{fig:vec}
The vector-meson ($V$) exchange terms.
From the first row to second row, $K^+$ and $M$ are exchanged.
}
\end{figure}

We consider gauge-invariant tree-level photo-production mechanisms 
($V^j_\alpha$) as follows:
minimal substitution to the lowest order chiral meson-baryon interaction
such as the Weinberg-Tomozawa terms (Fig.~\ref{fig:wt}) and the Born
terms (Fig.~\ref{fig:br1});
vector-meson exchange mechanisms (Fig.~\ref{fig:vec}).
Thus, for specifying each mechanism $\alpha$, we use the label for each figure
in Figs.~\ref{fig:wt}-\ref{fig:vec}  so that
$\alpha=$\,\ref{fig:wt}(a), \ref{fig:wt}(b), .., \ref{fig:vec}(k).
These photo-production mechanisms are expanded in terms of $1/M_B$,
and ${\cal O} (1)$ and ${\cal O} (1/M_B)$ terms are considered in our calculation.
Explicit expressions for $V^j_\alpha$, as well as our model Lagrangians
from which $V^j_\alpha$ are derived, are shown in
Appendix~\ref{app:lag} and \ref{app:ph-amp}.
Coupling constants contained in $V^j_\alpha$ of Figs.~\ref{fig:wt}-\ref{fig:vec} are fixed either by data 
(other than $\gamma p\to K^+\pi\Sigma$) if possible,
or by SU(3) relation if poorly constrained by data.
More details about the couplings are given in Appendix~\ref{app:lag}.

With the meson-exchange production mechanisms
and the subtraction constants ($a^j_\alpha(\mu)$ in Eq.~(\ref{eq:gpropdr}))
taken as the same as those in the chiral unitary amplitudes, 
we cannot reproduce the $\pi\Sigma$ line-shape data
for the $\gamma p\to K^+\pi\Sigma$ reaction
from the CLAS~\cite{Moriya:2013eb}.
Therefore, it is inevitable to introduce adjustable degrees of freedom
to fit the data.
Thus all of the meson-exchange mechanisms $V^j_\alpha$ are multiplied by a
common form factor of the following form:
\begin{eqnarray} 
\left(\Lambda^2\over \Lambda^2 + \bm k^{*2}\right)
\left(\Lambda^2\over \Lambda^2 + \bm k_j^{*2}\right) \ ,
\label{eq:ff}
\end{eqnarray}        
where $\bm k^*$ and $\bm k_j^*$ are respectively the momenta of $K^+$ and $M_j$ in
the center-of-mass frame of $M_jB_j$.
The cutoff $\Lambda$ will be used to fit the data.
In addition, we also consider
phenomenological contact terms that can simulate mechanisms not
explicitly considered, such as, in particular, $N^*$ and $Y^*$ excitation mechanisms.
We take couplings for the contact terms $W$-dependent ($W$ : total energy of the system),
and will be determined by fitting the $\gamma p\to K^+\pi\Sigma$ data~\cite{Moriya:2013eb}.
We consider three types of contact terms that 
are gauge-invariant at the tree-level, and are
couple to $K^+\bar KN$ 
and $K^+\pi\Sigma$ states of different charges, and thus 
we have 15 complex couplings at each $W$.
Expressions for the contact terms are presented in
Eqs.~(\ref{eq:c1})-(\ref{eq:c3}) in Appendix.
Also, for the mechanism index $\alpha$, we write 
$\alpha=c1,c2,c3$, as in 
Eqs.~(\ref{eq:c1})-(\ref{eq:c3}).
The form factor of Eq.~(\ref{eq:ff}) is not applied to the contact terms.

The subtraction constants
$a^j_\alpha(\mu)$ included in Eq.~(\ref{eq:gpropdr}) 
are also adjusted to fit the data, thereby changing the interference
pattern between different production mechanisms.
As already stated, 
$a^j_\alpha(\mu)$ can depend on the production mechanism
$\alpha=$\,\ref{fig:wt}(a), \ref{fig:wt}(b), .., \ref{fig:vec}(k), 
$c1,c2,c3$.
However, some of $\alpha$'s should have the same value for $a^j_\alpha(\mu)$.
Also, we do not want to have too many free parameters from the
subtraction constants, because it will complicate fitting the data. 
Thus we classify the production mechanisms into several groups, and each
group has its own real subtraction constant.
In grouping, we try to classify important mechanisms into different
groups so that we have effective freedom in fitting.
In TABLE~\ref{tab:sub}, we show the classification of the mechanisms
into 11 groups labeled by A,B,...,K.
\begin{table}
\caption{\label{tab:sub} Classification of production mechanisms.
The meson-exchange mechanisms shown in Figs.~\ref{fig:wt}-\ref{fig:vec}
and three contact terms $c1$-$c3$
are classified into 11 groups labeled by A,B,...,K. Each group has its own
 subtraction constants $a^j_{\alpha'}(\mu)$ in Eq.~(\ref{eq:gpropdr}).
}
\renewcommand{\arraystretch}{1.2}
\tabcolsep=3.4mm
\begin{tabular}{c|ccccc}\hline
&A&B&C&D&E\\\hline
$\alpha$& 
\ref{fig:br1}(a)-(d),\ref{fig:vec}(a),$c1,c2,c3$&
 \ref{fig:wt}(a),(c),(e)&
 \ref{fig:wt}(d)&
 \ref{fig:wt}(b)&
\ref{fig:br1}(h)-(j)
\\\hline
\end{tabular}

\vspace*{3mm}
\begin{tabular}{c|cccccc}\hline
&F&G&H&I&J&K\\\hline
$\alpha$& 
\ref{fig:br1}(k)-(n)&
\ref{fig:br1}(e),(g)&
\ref{fig:br1}(f)&
\ref{fig:vec}(b)-(d)&
\ref{fig:vec}(e)-(h)&
\ref{fig:vec}(i)-(k)
\\\hline
\end{tabular}
\end{table}
The subtraction constant for each group is denoted by
$a^j_{\alpha'}(\mu)$ where 
${\alpha'}$ refers to one of the groups, A,B,...,K.
Then $a^j_{\alpha'}(\mu)$
will be used to fit the data~\cite{Moriya:2013eb}.
It is noted that we do not adjust the subtraction constants in the
chiral unitary amplitudes in the fit.
The subtraction constants we adjusted are all for the first loop of the
rescattering, and for the renormalization of the production mechanism.

For the number of data points to
be fitted,
we now have rather many free parameters
most of which are from contact terms.
We find that this amount of degrees of freedom is necessary to obtain a
reasonable fit to the data.
This situation can be understood,
considering that we do not explicitly consider short-range mechanisms
(baryon resonances, coupled-channel effects)
that will play a substantial role here. 
Because it will be a very difficult task to identify and/or fix each of the short-range
mechanisms, we develop the production model in a practical manner as
discussed above.
Of course, our method could bring a model-dependence of $\Lambda(1405)$
pole(s) extracted from the data.
The model-dependence of $\Lambda(1405)$ pole(s) 
must be assessed by analyzing the data with
different form factors and/or contact terms.
This will be a future work. 

\section{Result}
\label{sec:result}

\subsection{Fitting data}
Before presenting our results, we comment on the calculated quantity to
be fitted to the line-shape data from CLAS~\cite{Moriya:2013eb}.
In the data analysis done by CLAS in Ref.~\cite{Moriya:2013eb}, enhanced
events due to the $K^*$ peak in the $\pi K^+$ invariant mass spectrum
has been subtracted. 
In our model, a mechanism of Fig.~\ref{fig:vec}(i) without rescattering
can create the $K^*$ peak.
Thus we fit the data with the following modified differential cross
section:
\begin{eqnarray}
\label{eq:xs-mod}
{d\sigma \over d M_{\pi\Sigma}} ({\rm full})
-{d\sigma \over d M_{\pi\Sigma}} (V^{\pi\Sigma}_{\ref{fig:vec}({\rm i})})
\ ,
\end{eqnarray}
where ${d\sigma \over d M_{\pi\Sigma}} ({\rm full})$ contains all of
the meson-exchange mechanisms and contact terms followed by the
rescattering as discussed in
Sec.~\ref{sec:ppm}, and is calculated using Eq.~(\ref{eq:ls}).
Meanwhile, the second term 
${d\sigma \over d M_{\pi\Sigma}} (V^{\pi\Sigma}_{\ref{fig:vec}({\rm i})})$
contains only the tree-level mechanism of Fig.~\ref{fig:vec}(i).
The subtraction in Eq.~(\ref{eq:xs-mod}) is done at the cross section
level, and the interference between the mechanism of
Fig.~\ref{fig:vec}(i) and others is kept to be consistent with the
analysis of Ref.~\cite{Moriya:2013eb}.

We present all numerical values for the fitting parameters (the cutoff from
the form factor, the subtraction constants, and the complex couplings
from the contact terms) in Appendix~\ref{app:fit-coup}.

\subsection{Line-shape results}

\begin{figure}
\includegraphics[clip,width=0.45\textwidth]{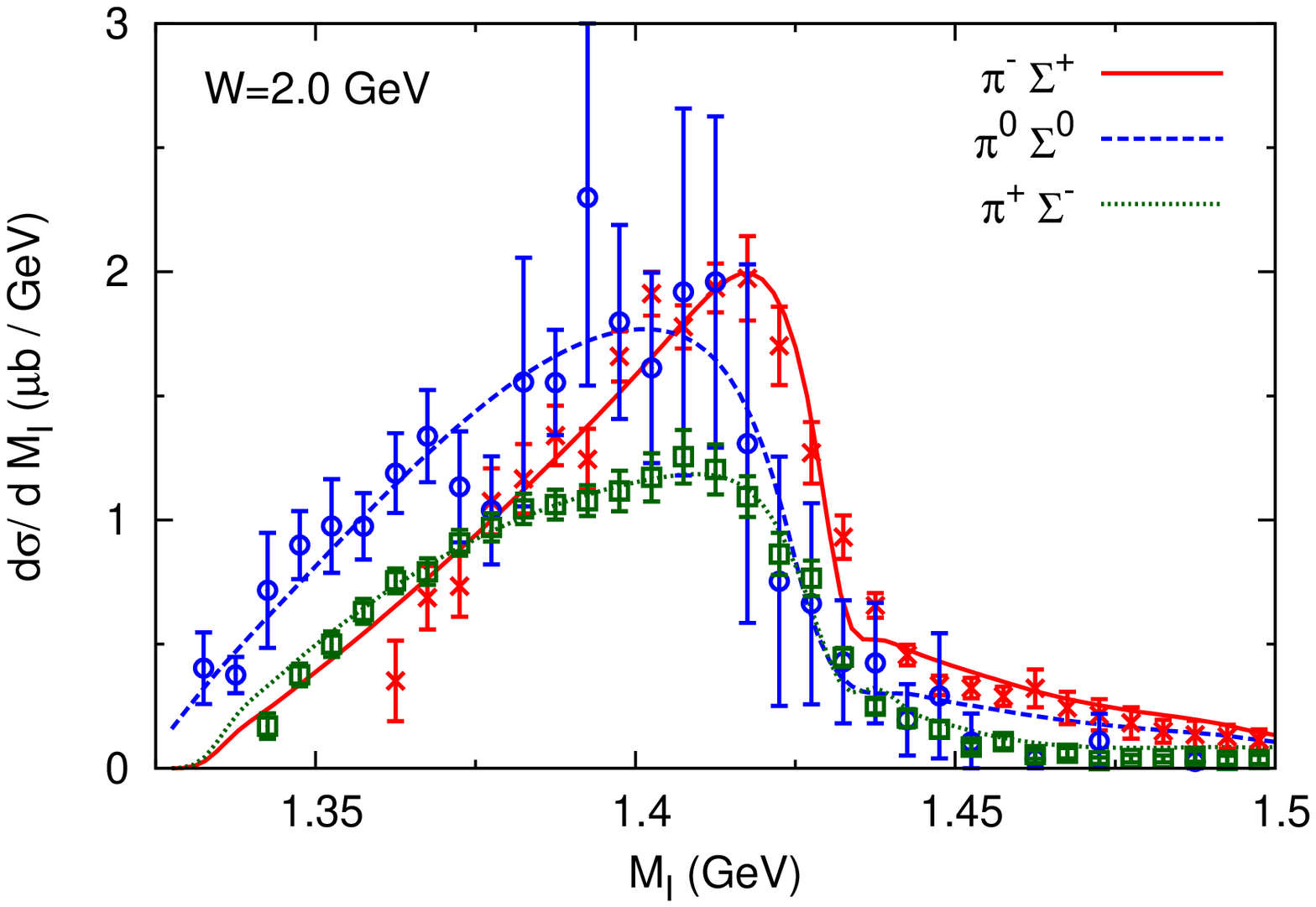}
\includegraphics[clip,width=0.45\textwidth]{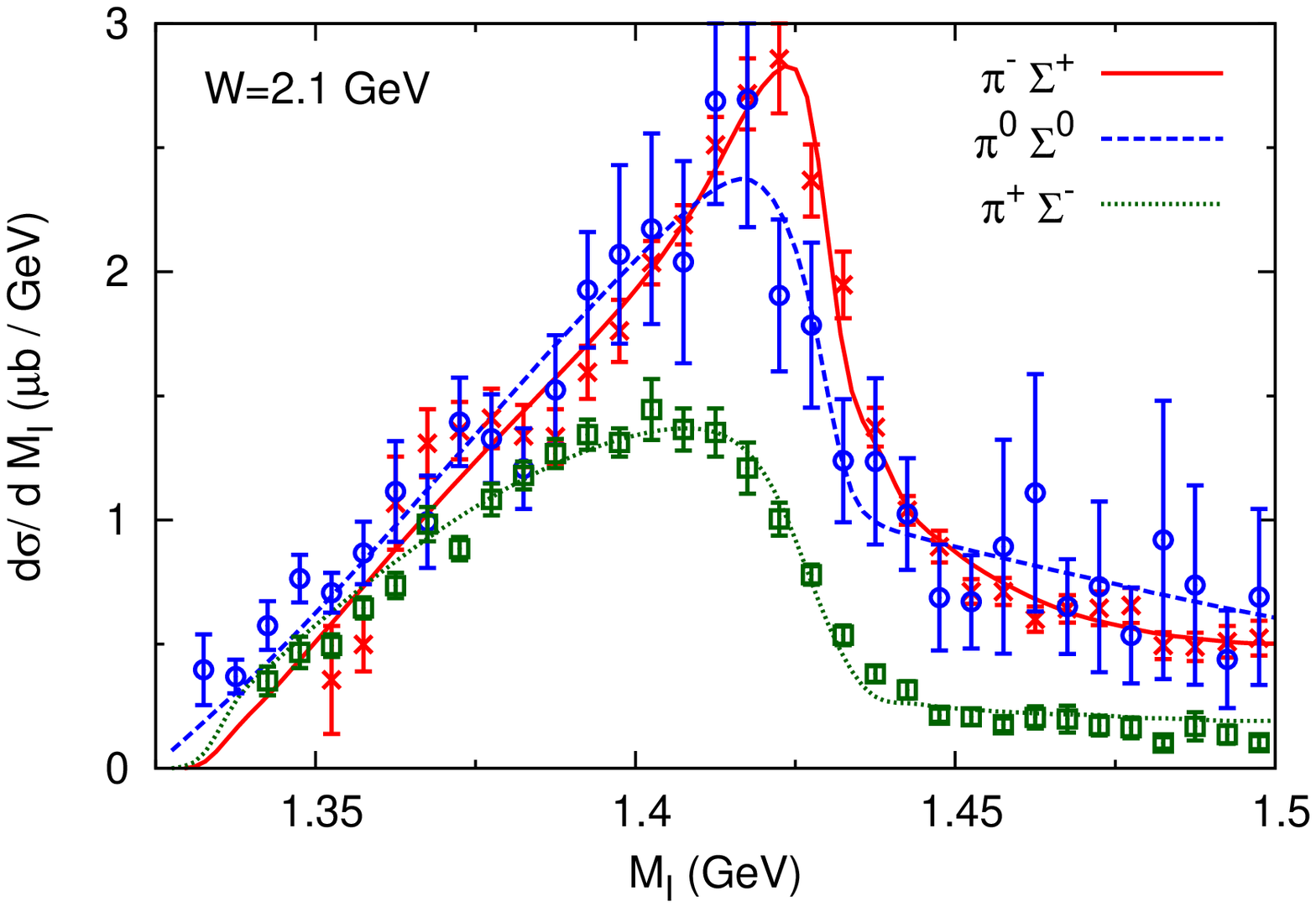}
\includegraphics[clip,width=0.45\textwidth]{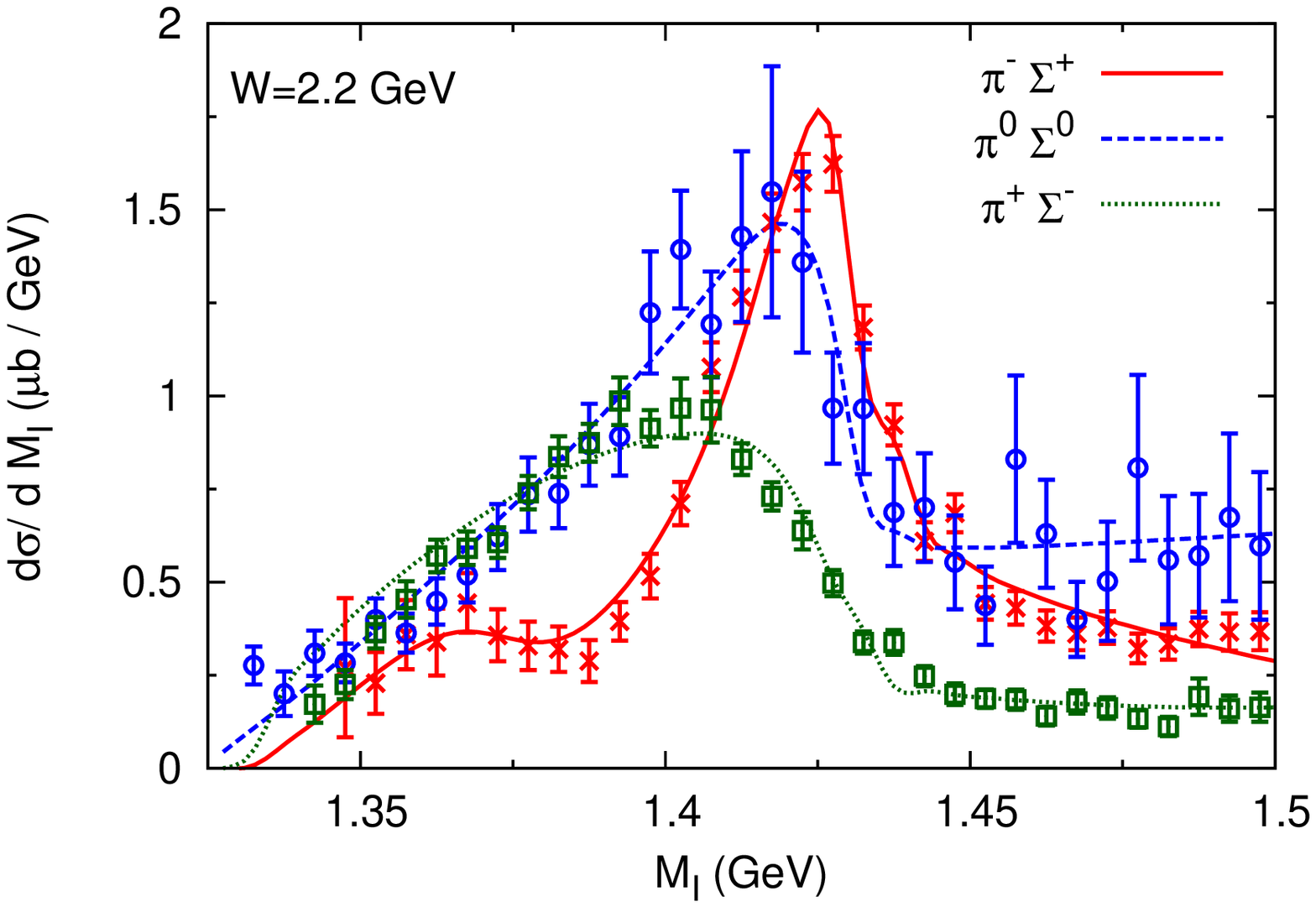}
\includegraphics[clip,width=0.45\textwidth]{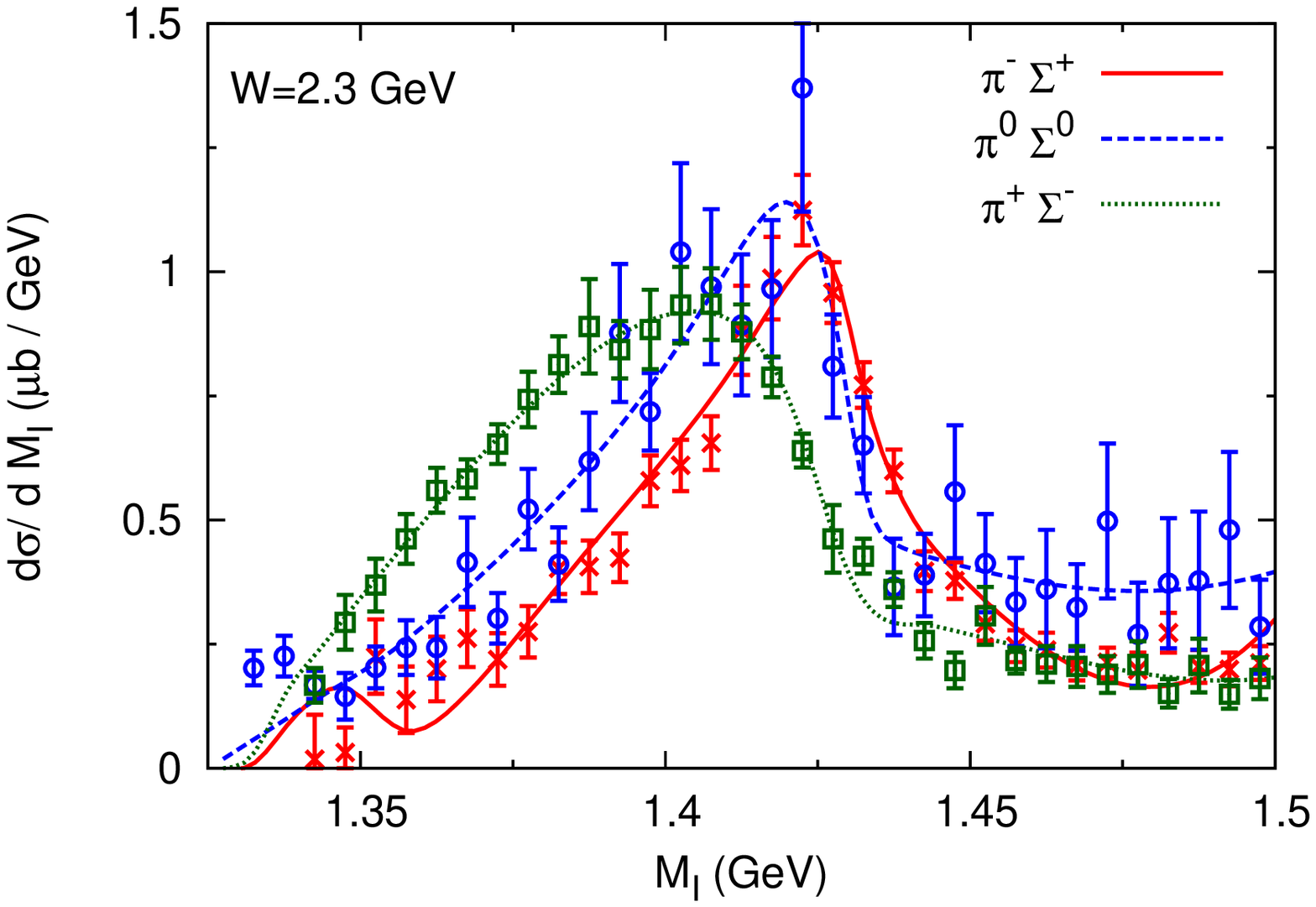}
\caption{\label{fig:2gev} (Color online)
Comparison of $\pi\Sigma$ line-shapes from our model with
 data~\cite{Moriya:2013eb} at $W=2.0$, 2.1, 2.2 and 2.3~GeV.
Symbols for the data are cross (red) for $\pi^-\Sigma^+$,
circle (blue) for $\pi^0\Sigma^0$, and
square (green) for $\pi^+\Sigma^-$.
}
\end{figure}

Our results, after the fit, are presented in Fig.~\ref{fig:2gev} where
the CLAS data are also shown for comparison. 
We fitted the data at $W=2.0$, 2.1, 2.2 and 2.3~GeV.
As seen in the figure, our model fits the data very well for all three
different charge states of $\pi\Sigma$.

\begin{figure}
\includegraphics[clip,width=0.45\textwidth]{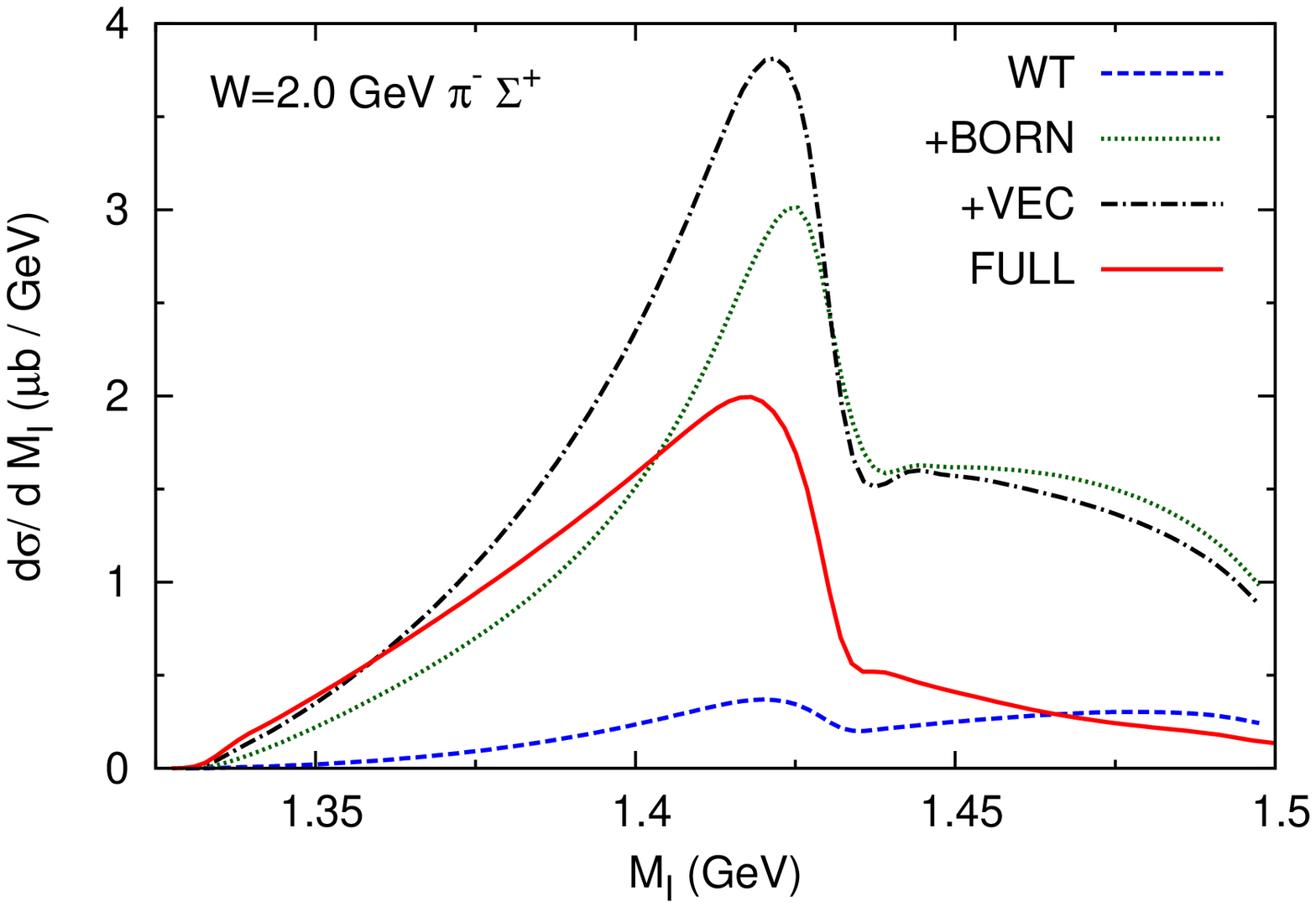}
\includegraphics[clip,width=0.45\textwidth]{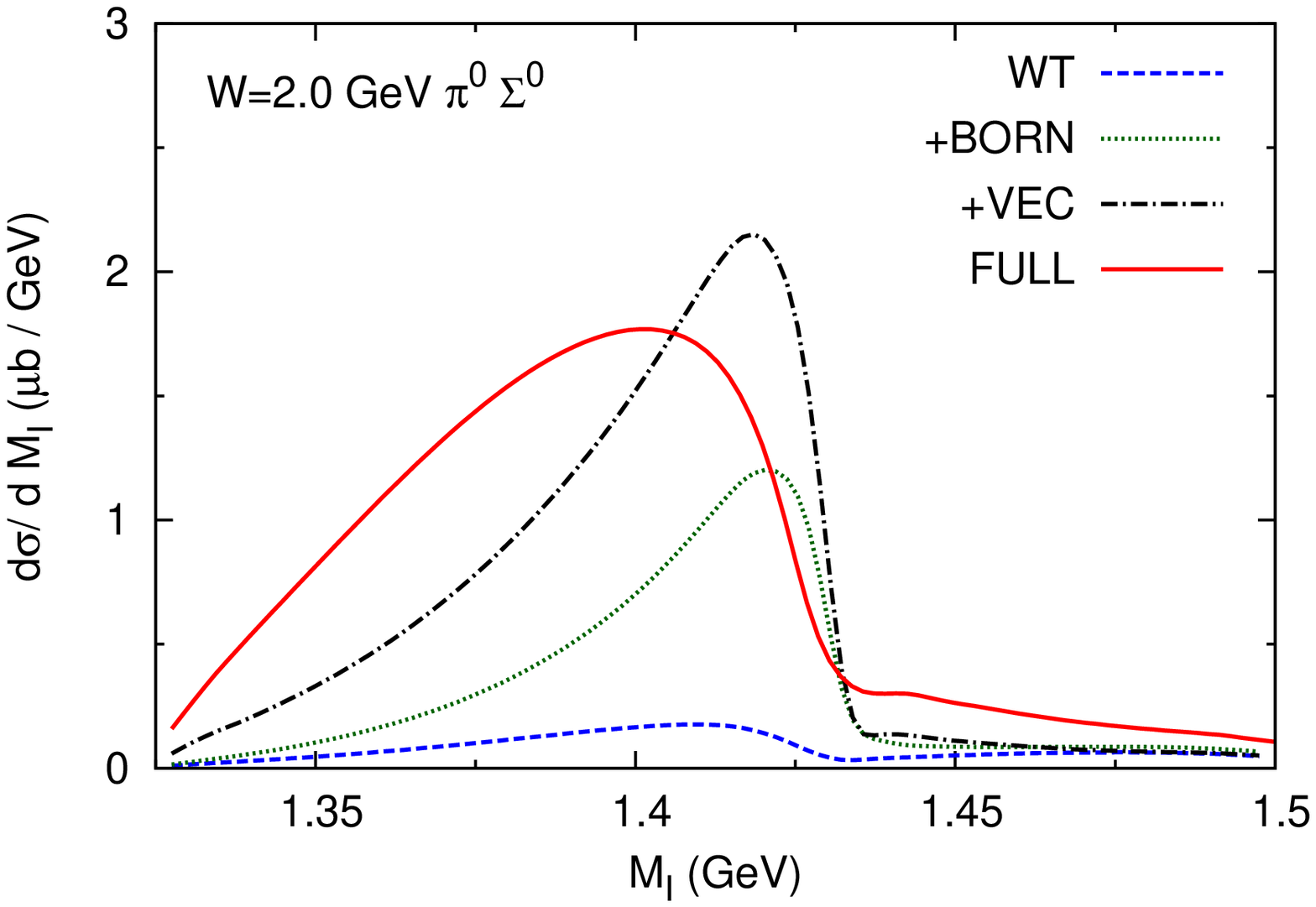}
\includegraphics[clip,width=0.45\textwidth]{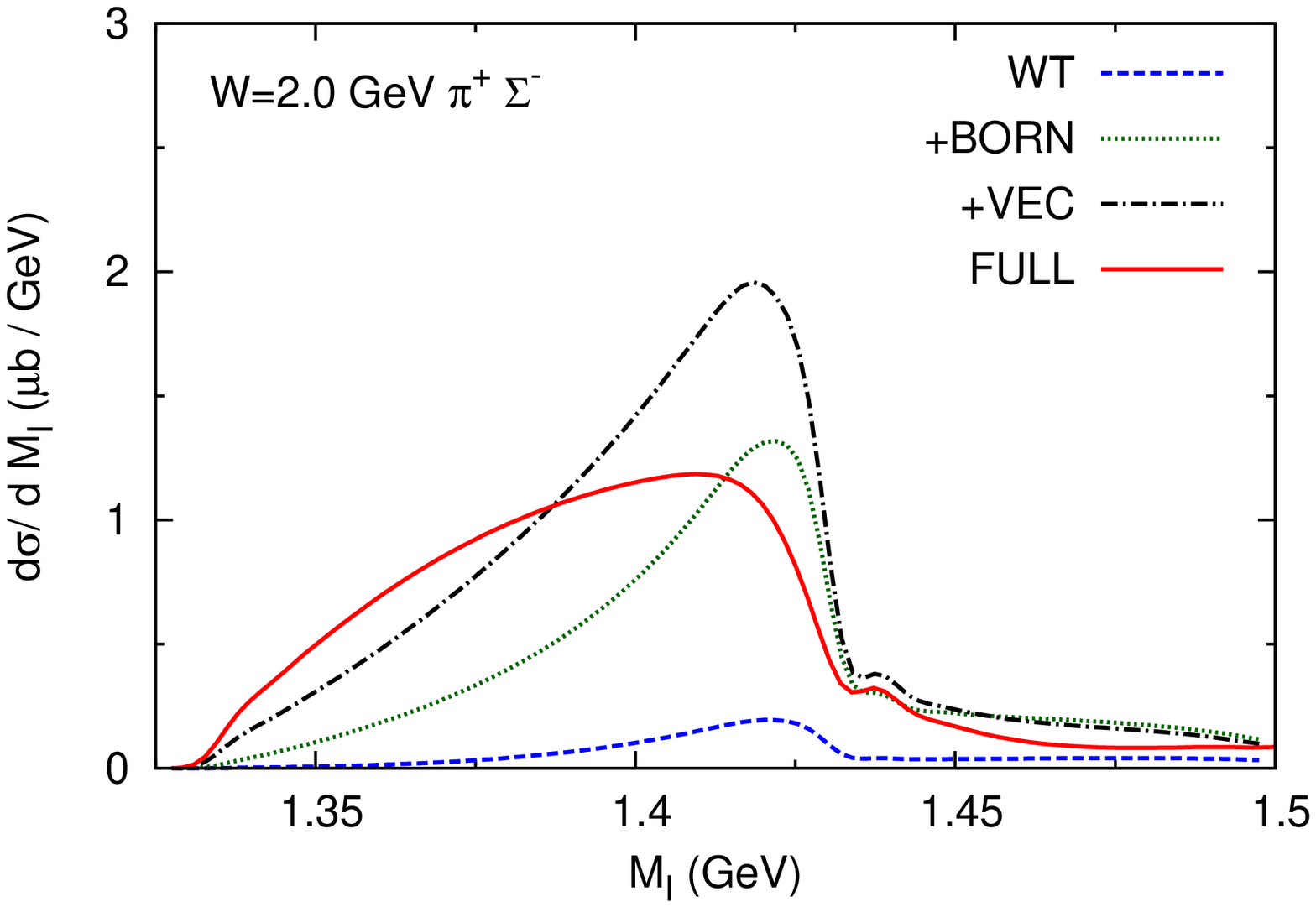}
\caption{\label{fig:mec} (Color online)
Contribution of each production mechanism.
Contribution from the gauged Weinberg-Tomozawa terms (Fig.~\ref{fig:wt})
is given by the blue dashed lines.
Contribution that additionally includes the gauged Born terms (Fig.~\ref{fig:br1})
is given by the green dotted lines.
Contribution that further includes the vector-meson exchange terms
 (Fig.~\ref{fig:vec}) is given by the black dash-dotted lines.
Finally, the full result, including the contact terms, 
is shown by the red solid lines.
}
\end{figure}

\begin{figure}
\includegraphics[clip,width=0.45\textwidth]{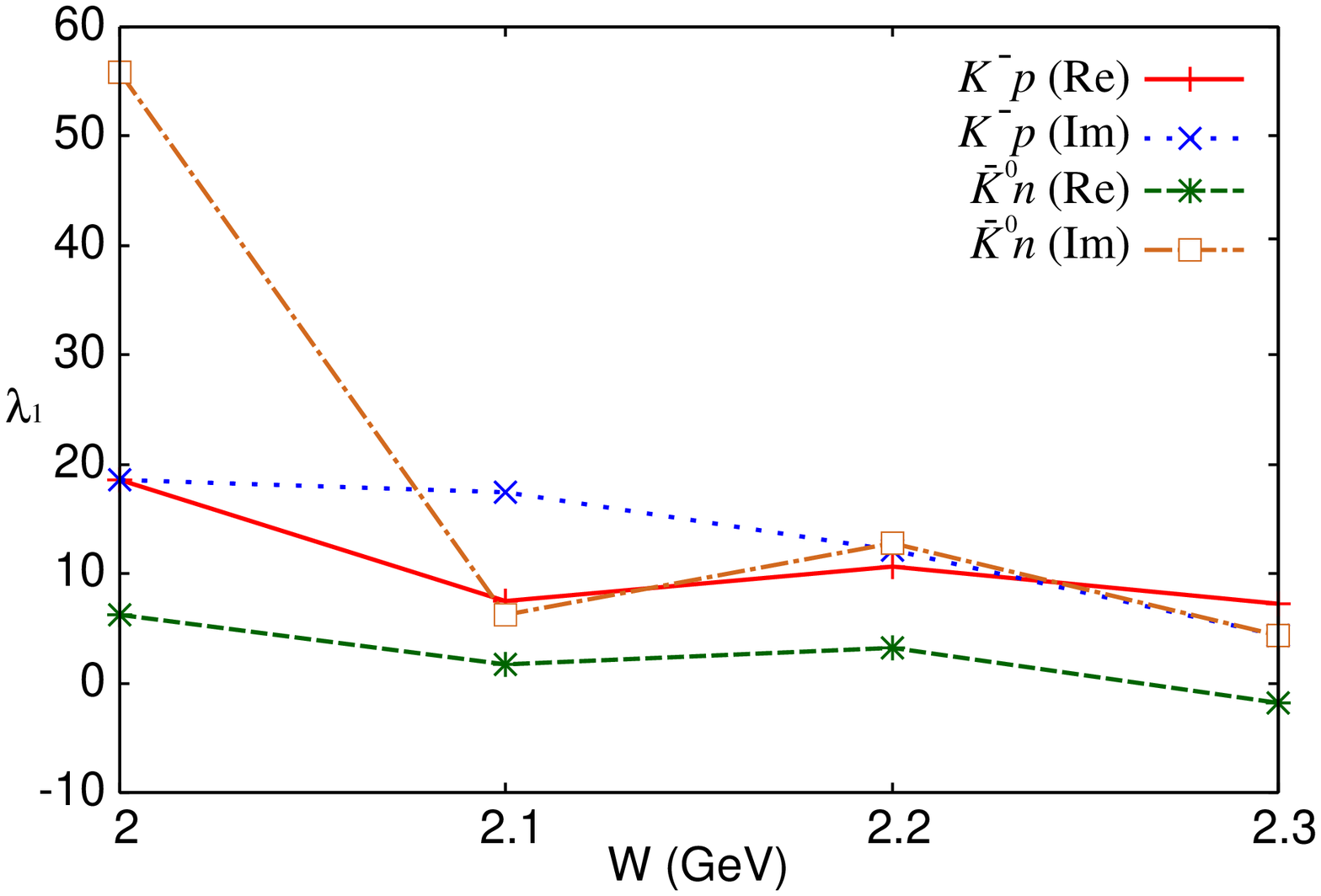}
\hspace{3mm}
\includegraphics[clip,width=0.45\textwidth]{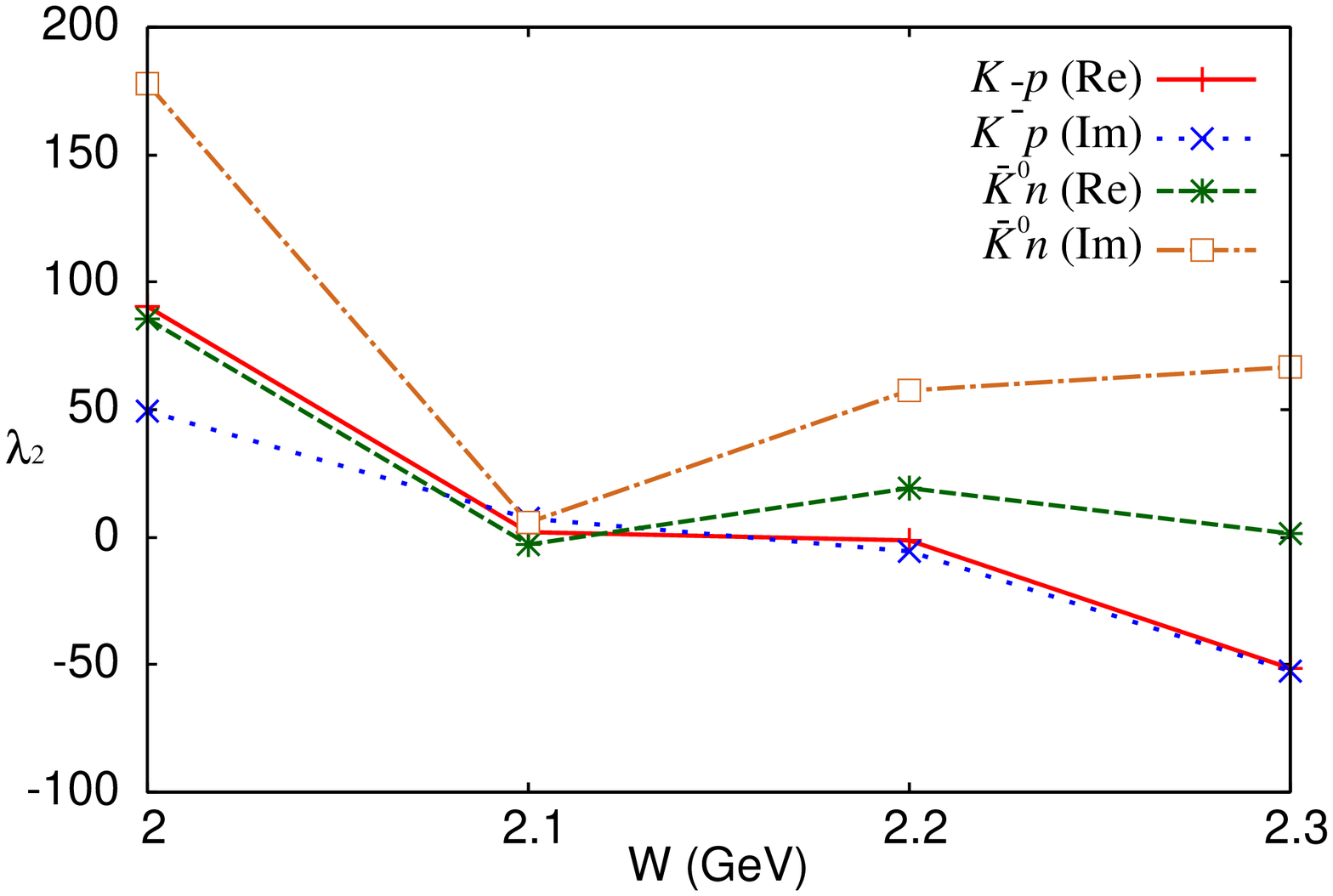}
\caption{\label{fig:contact} (Color online)
The $W$-dependence of the contact couplings, 
$\lambda^j_1$ (left) and $\lambda^j_2$ (right) for $j=K^- p$ and $\bar{K}^0n$.
The contact terms are defined in Eqs.~(\ref{eq:c1})-(\ref{eq:c3}).
The real and imaginary parts of the couplings are indicated by (Re) and
 (Im), respectively.
}
\end{figure}

It is interesting to break down the line-shapes into 
contributions from different mechanisms, as shown in Fig.~\ref{fig:mec}.
As seen in the figures, different mechanisms gives significant
contributions that interfere with each other.
We find that the contributions from the gauged Weinberg-Tomozawa terms (Fig.~\ref{fig:wt})
are rather small. 
In fact, a diagram such as Fig.~\ref{fig:wt}(e) gives a contribution
comparable to those from the gauged Born mechanisms (Fig.~\ref{fig:br1}).
However, as a result of a destructive interference, the net contribution
from the gauged Weinberg-Tomozawa terms is rather small.
This destructive interference is not necessarily a result of the gauge
invariance. 
Actually, a dominant term in Fig.~\ref{fig:wt}(e) is gauge invariant
itself.
Rather, fitting the data have fixed relevant subtraction constants 
so that the diagrams in Fig.~\ref{fig:wt} with the rescattering
cancel out each other.
We find relatively large contributions from mechanisms of 
Fig.~\ref{fig:wt}(e), \ref{fig:br1}(b), \ref{fig:br1}(e),
\ref{fig:br1}(i), \ref{fig:br1}(l), 
\ref{fig:vec}(b), \ref{fig:vec}(f),
that have two propagators rather than three in the other mechanisms;
the propagators tend to suppress the contributions of the mechanisms.
Meanwhile, the contact terms, which simulate short-range dynamics,
also give a large contribution to bring the theoretical calculation
into agreement with the data. 
As seen in TABLE~\ref{tab:fit-param}, the contact terms have a rather
strong coupling to the $\bar K N$ channels as a result of the fit.
One may find in TABLE~\ref{tab:fit-param}
that the $W$-dependence of the
contact couplings is rather irregular, and is not well under control. 
However, we note that the contact terms can have a resonant behavior. 
Also, in Fig.~\ref{fig:contact}, 
we show the $W$-dependence of the most
important contact couplings, 
$\lambda^j_1$ and $\lambda^j_2$ for $j=K^- p$ and $\bar{K}^0n$.
From the figure, it is hard to judge if the behavior of the
couplings is out of control. 
As will be discussed later, however,
we would be able to put them under better control 
if we fit not only the line-shape
data but also other observables such as angular distributions.
This will be a future work.
Finally, we mention that coupled-channels effects are mostly from 
the $\bar K N$ and $\pi\Sigma$ channels.

\begin{figure}
\includegraphics[clip,width=0.45\textwidth]{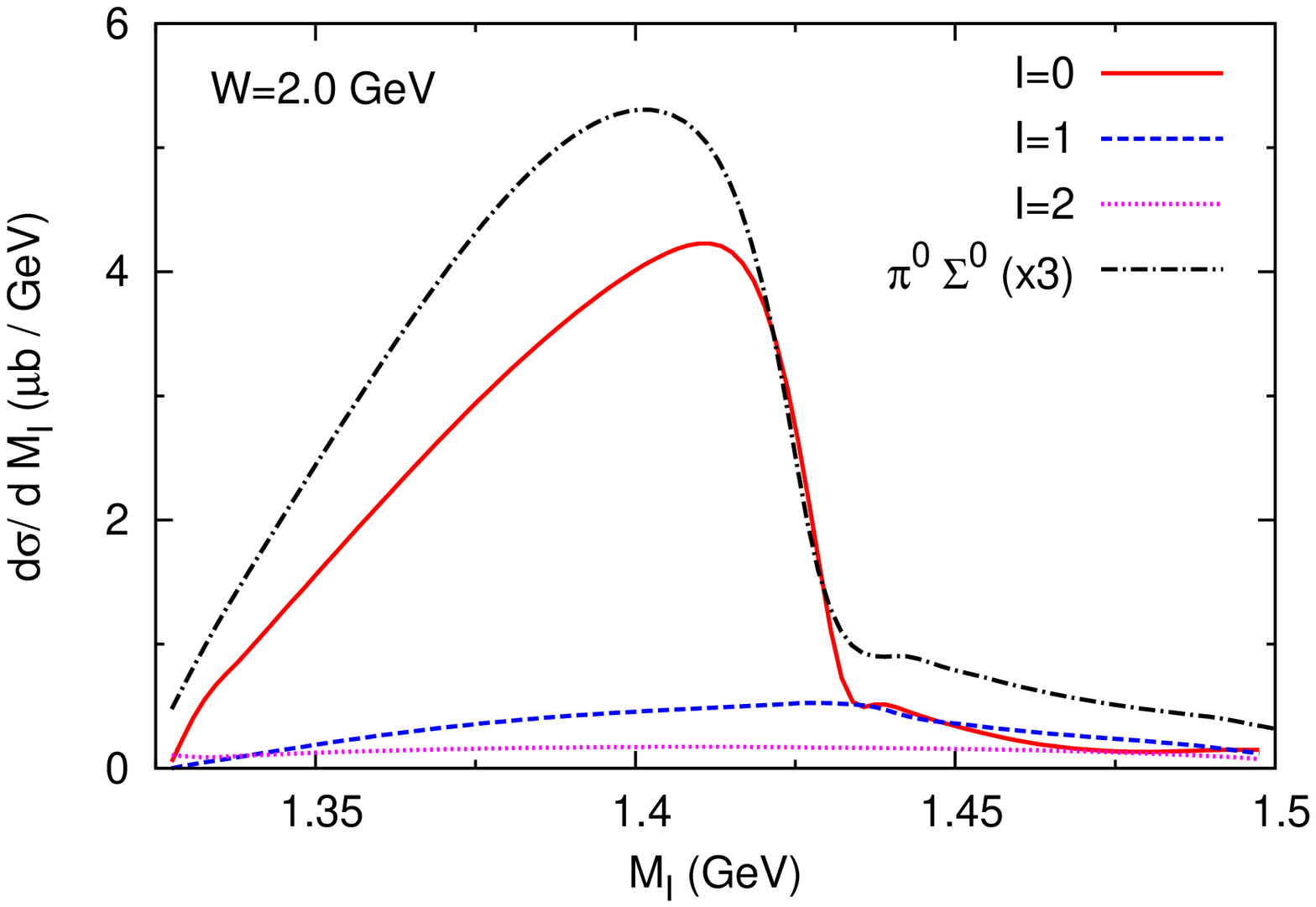}
\includegraphics[clip,width=0.45\textwidth]{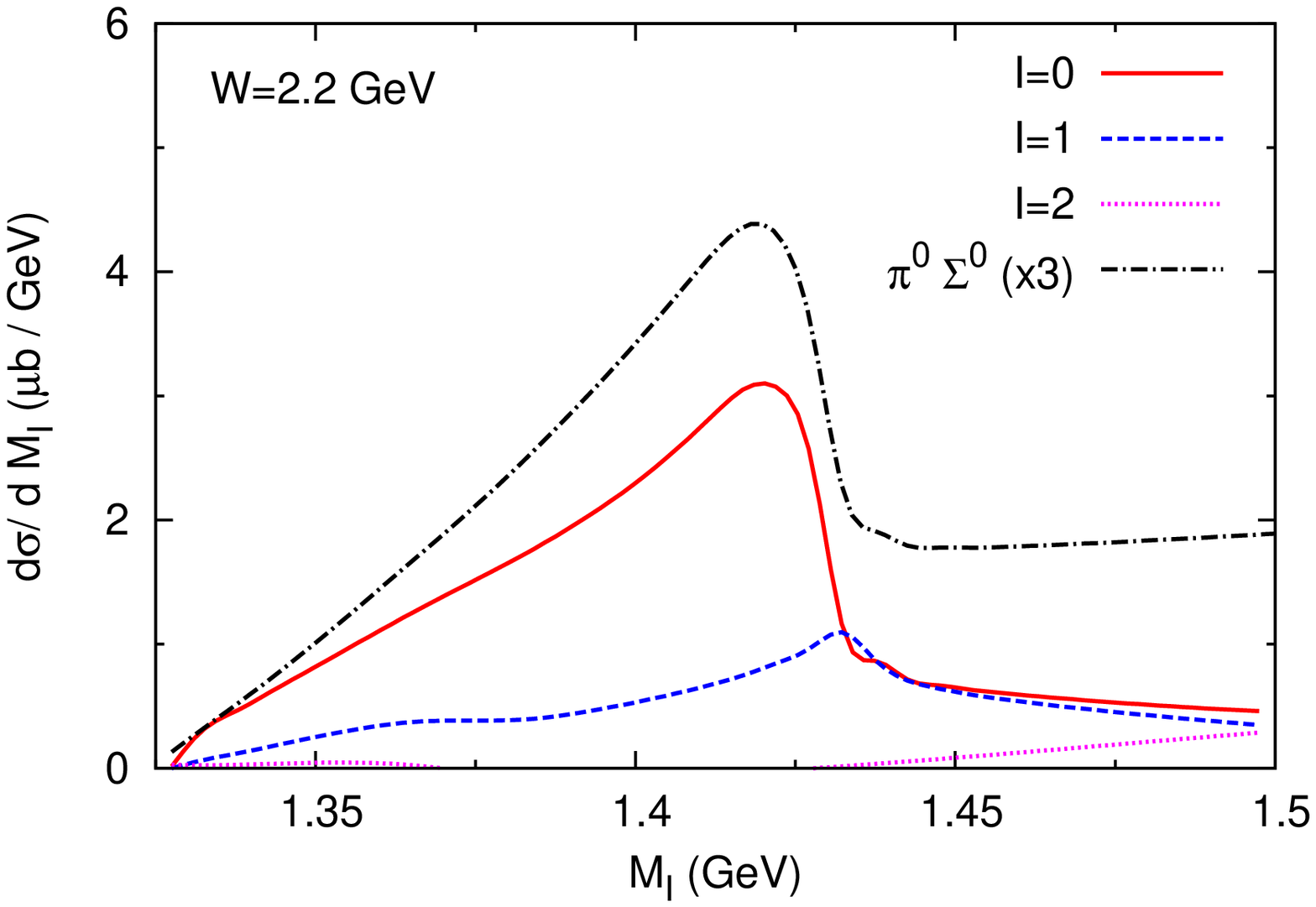}
\caption{\label{fig:iso-2gev} (Color online)
Isospin decomposition of $\pi\Sigma$ line-shapes. Contributions from the
 total isospin state $I$ are shown, along with the line-shape of
 $\pi^0\Sigma^0$ multiplied by 3.
}
\end{figure}

The difference in the line-shape
between different charge states observed in Fig.~\ref{fig:2gev}
is a result of the interference between different isospin states.
The $\pi\Sigma$ has three isospin states ($I=0,1,2$), and they are
separately shown in Fig.~\ref{fig:iso-2gev}.
A dominant contribution is from the $I=0$ state as expected due to the
$\Lambda(1405)$ peak. 
The higher mass pole at $1426- 16i$ MeV, that creates the prominent bump
in the line-shapes, seems to play more important role than the lower
mass pole. 
This is because the production mechanisms in our model generate $\bar K N$ more strongly
than $\pi\Sigma$, and the final state interaction induces 
$\bar K N\to \pi\Sigma$.
As shown in the previous study~\cite{Jido:2003cb}, the higher mass pole 
couples to the $\bar K N$ channel more strongly than the lower mass pole
does. 
The $I=1$ state gives a smaller contribution, still plays an important
role to generate the charge dependence of the $\pi\Sigma$ line-shapes. 
As the energy $W$ increases, the $I=1$ contribution is larger.
The $I=2$ state contribution is even more smaller, but still
unnegligible.
To see this point, we show in Fig.~\ref{fig:iso-2gev} 
the $\pi^0\Sigma^0$ line-shape multiplied by 3.
The difference between this and the $I=0$ line-shape is the effect of
the interference between the $I=0$ and $I=2$ states. 
We can see that the interference with the $I=2$ state even changes slightly the peak
position of the $\pi^0\Sigma^0$ line-shape.

\begin{figure}
\includegraphics[clip,width=0.45\textwidth]{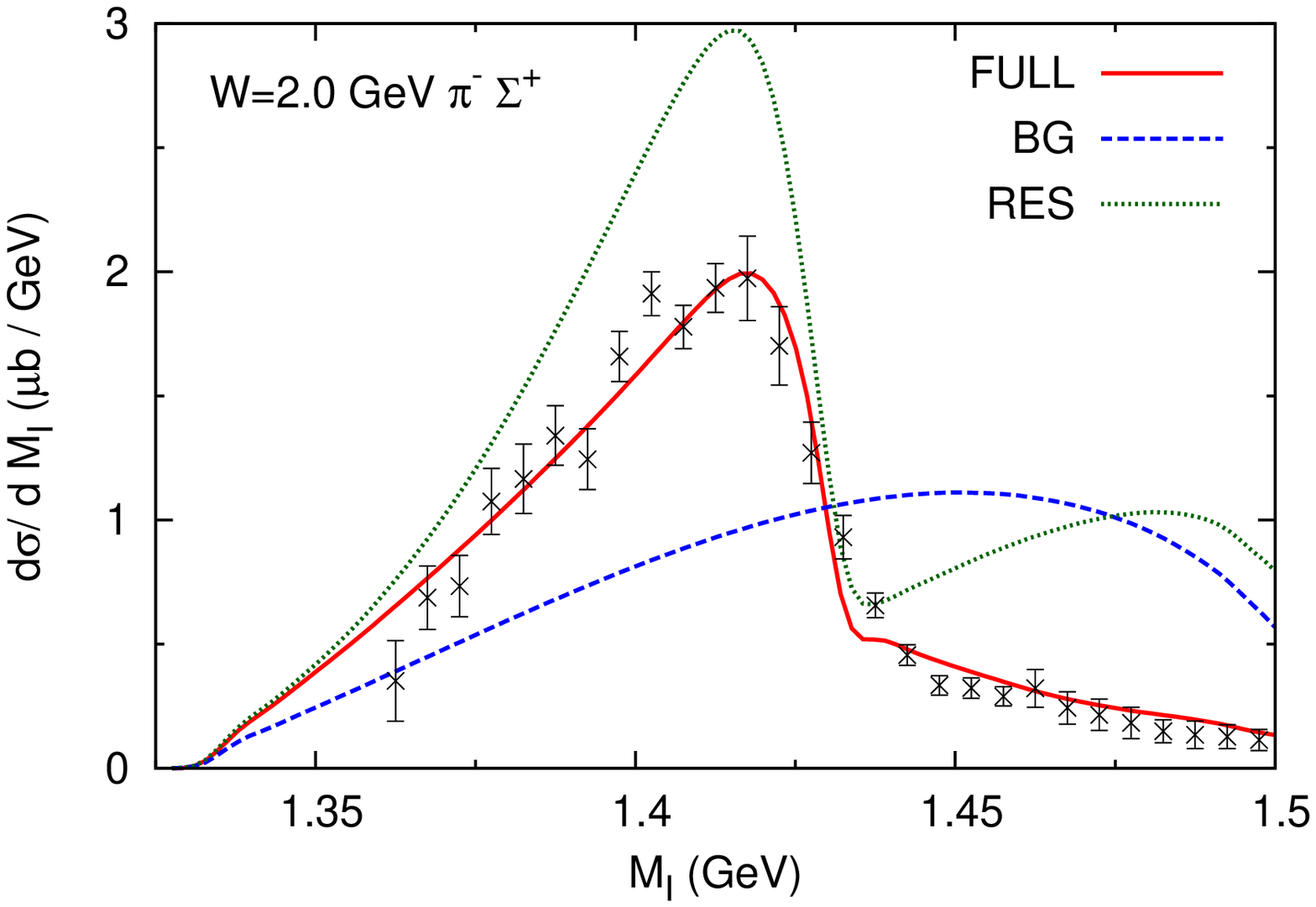}
\includegraphics[clip,width=0.45\textwidth]{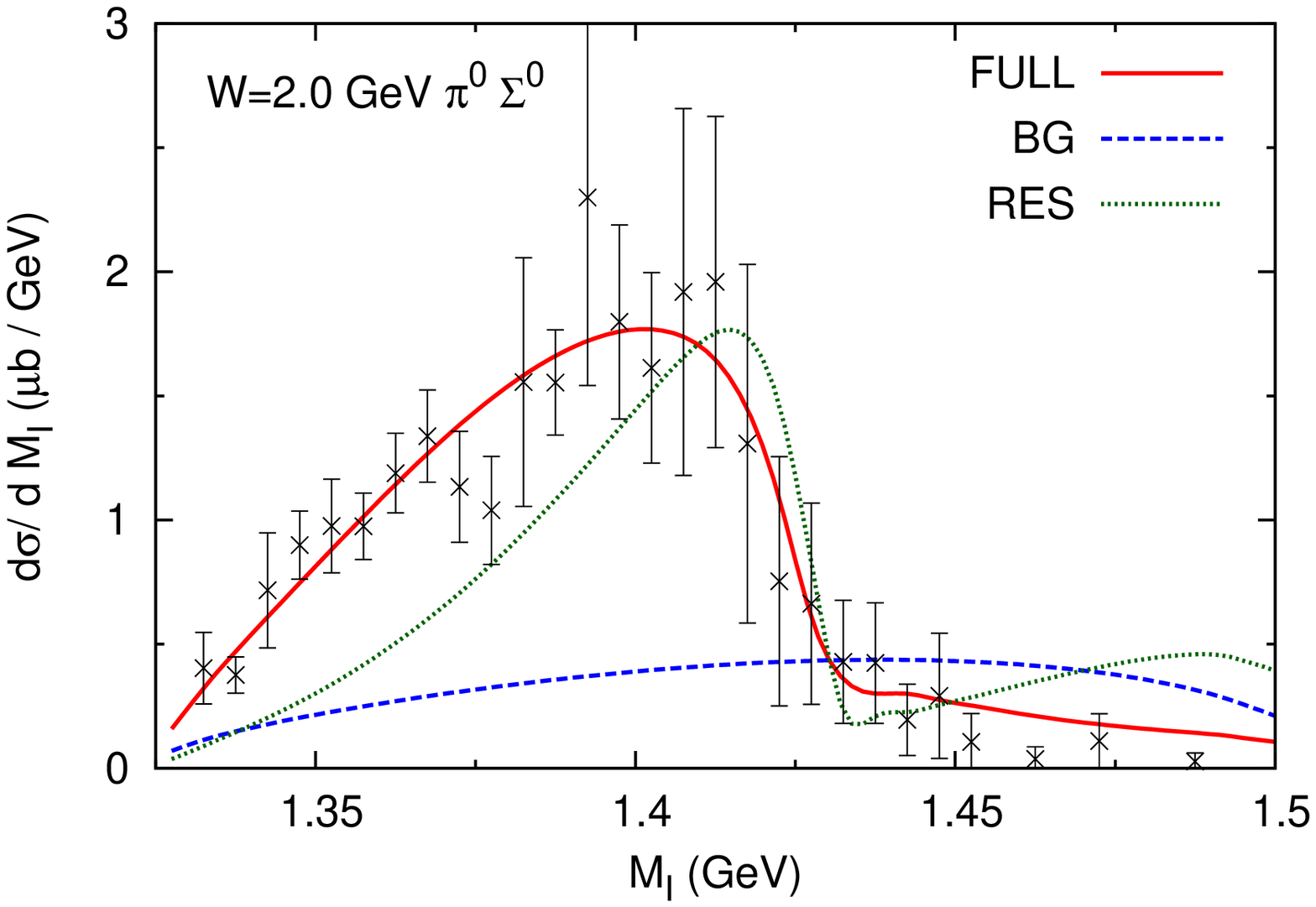}
\includegraphics[clip,width=0.45\textwidth]{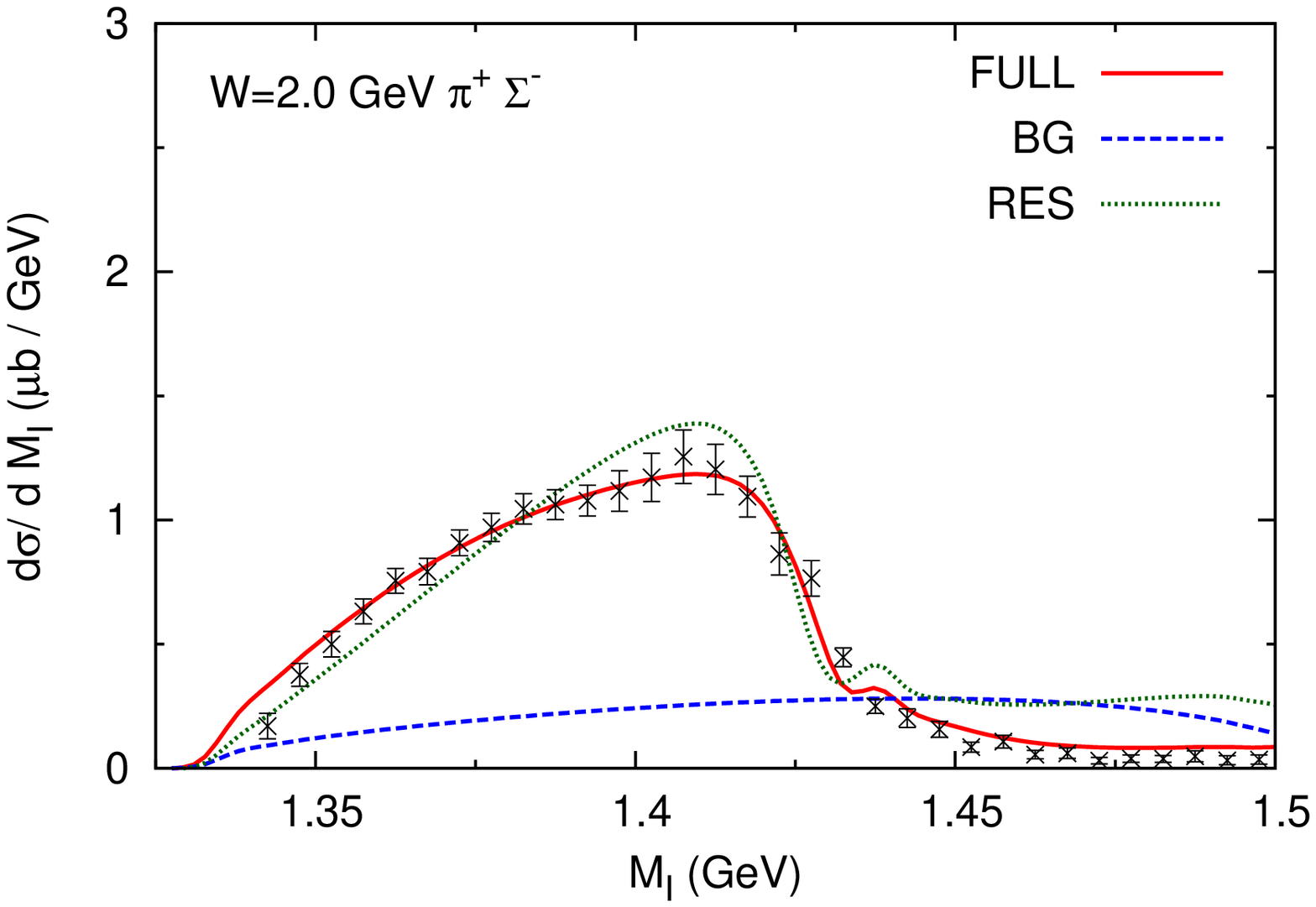}
\caption{\label{fig:bg-res} (Color online)
Contribution of background (BG)
and resonant (RES) terms.
Contribution of 
the background contribution is from first term of Eq.~(\ref{eq:tamp})
while the resonant contribution is from second term of Eq.~(\ref{eq:tamp}).
Coherent sum of 'BG' and 'RES' is given by 'FULL', and is compared with
 data~\cite{Moriya:2013eb}.
}
\end{figure}

The different peak positions for differently charged $\pi\Sigma$ states
seen in Fig.~\ref{fig:2gev} can be also understood as 
a result of the interference between a resonant and a background parts. 
To see this, 
it is useful to decompose the amplitude into
the resonant (second term of Eq.~(\ref{eq:tamp}))
and background (first term of Eq.~(\ref{eq:tamp})) parts.
Each of the contributions to $\gamma p\to K^+\pi\Sigma$ at $W=2.0$~GeV
is shown in Fig.~\ref{fig:bg-res}.
Interestingly, the background terms give smooth and significant
contributions.
Although the peak structures are due to the resonant contributions, 
the background can shift the positions of the peaks, particularly the
peak of the $\pi^0\Sigma^0$ line-shape.
After all, the resonant contributions give the peaks at almost the same
position, $M_{\pi\Sigma}\sim 1.42$~GeV, for all the differently charged $\pi\Sigma$ states. 
Thus it seems that one of the $\Lambda(1405)$ poles at
$1426- 16i$ MeV plays a dominant role in the line-shapes.
We will look into this observation in the next subsection.

\subsection{Single Breit-Wigner model}

So far, the excitation of $\Lambda(1405)$ is described by the chiral
unitary model, and $\Lambda(1405)$ has the double-pole structure.
However, as seen above, only the higher mass pole seems to give the
dominant contribution.
Thus, it is interesting to see if a single Breit-Wigner can simulate
the photo-induced $\Lambda(1405)$ excitation.
For this purpose, 
we use a model in which 
the rescattering amplitude,
$T^{jj'}_{JL}$ in Eq.~(\ref{eq:pw}), is given by, instead of 
the chiral unitary amplitude,
a single Breit-Wigner
function in $(J,L)=(1/2,0)$ and isospin zero partial wave;
the other rescattering partial waves amplitudes are set to zero.
Here we assume that the rescattering amplitude couples to only $\bar K N$
and $\pi\Sigma$ channels. Thus we have
\begin{eqnarray}
\label{eq:bw}
T^{jj'}_{1/2,0}(s) = {C^{BW}_{jj'}\over \sqrt{s}-M_{BW}+i
{\Gamma_{BW}\over 2}} \ ,
\end{eqnarray}
where $M_{BW}$ and $\Gamma_{BW}$ are Breit-Wigner mass and width,
respectively.
A symbol $C^{BW}_{jj'}$ is a complex coupling strength, and will be
fitted to the data, along with the Breit-Wigner mass, width and other fitting parameters.
Because of being isospin zero, 
$C^{BW}_{jj'}$ have three independent complex values that we denote
$C^{BW}_{\bar KN, \bar KN}$,
$C^{BW}_{\bar KN, \pi\Sigma}$, and
$C^{BW}_{\pi\Sigma, \pi\Sigma}$:
$C^{BW}_{\bar KN, \bar KN}$ is for $(j,j')=(1\,{\rm or}\,2, 1\,{\rm or}\,2)$;
$C^{BW}_{\bar KN, \pi\Sigma}$ is for
$(j,j')\, {\rm or}\, (j',j)=(4\,{\rm or}\,7\,{\rm or}\,8, 1\,{\rm or}\,2)$;
$C^{BW}_{\pi\Sigma, \pi\Sigma}$ is for 
$(j,j')\, =(4\,{\rm or}\,7\,{\rm or}\,8, 4\,{\rm or}\,7\,{\rm or}\,8)$.
We have defined the index $j$ at the beginning of Sec.~\ref{sec:ppm}.
Our Breit-Wigner form of Eq.~(\ref{eq:bw}) is more relaxed than the
conventional one in which the coupling strengths are related to the
width by 
$\Gamma_{BW} = \sum_j \delta(\sqrt{s}-\sqrt{s_j})  C^{BW}_{jj}$, 
where the summation is taken over both channels and their particles'
phase-space.
In this way, we can simulate non-resonant effects that are not
considered explicitly in the rescattering amplitude.
The single Breit-Wigner model is fitted to the data and is shown with
the data in Fig.~\ref{fig:bw}.
The fitted parameters are presented in tables in Appendix~\ref{app:fit-coup}.
\begin{figure}
\includegraphics[clip,width=0.45\textwidth]{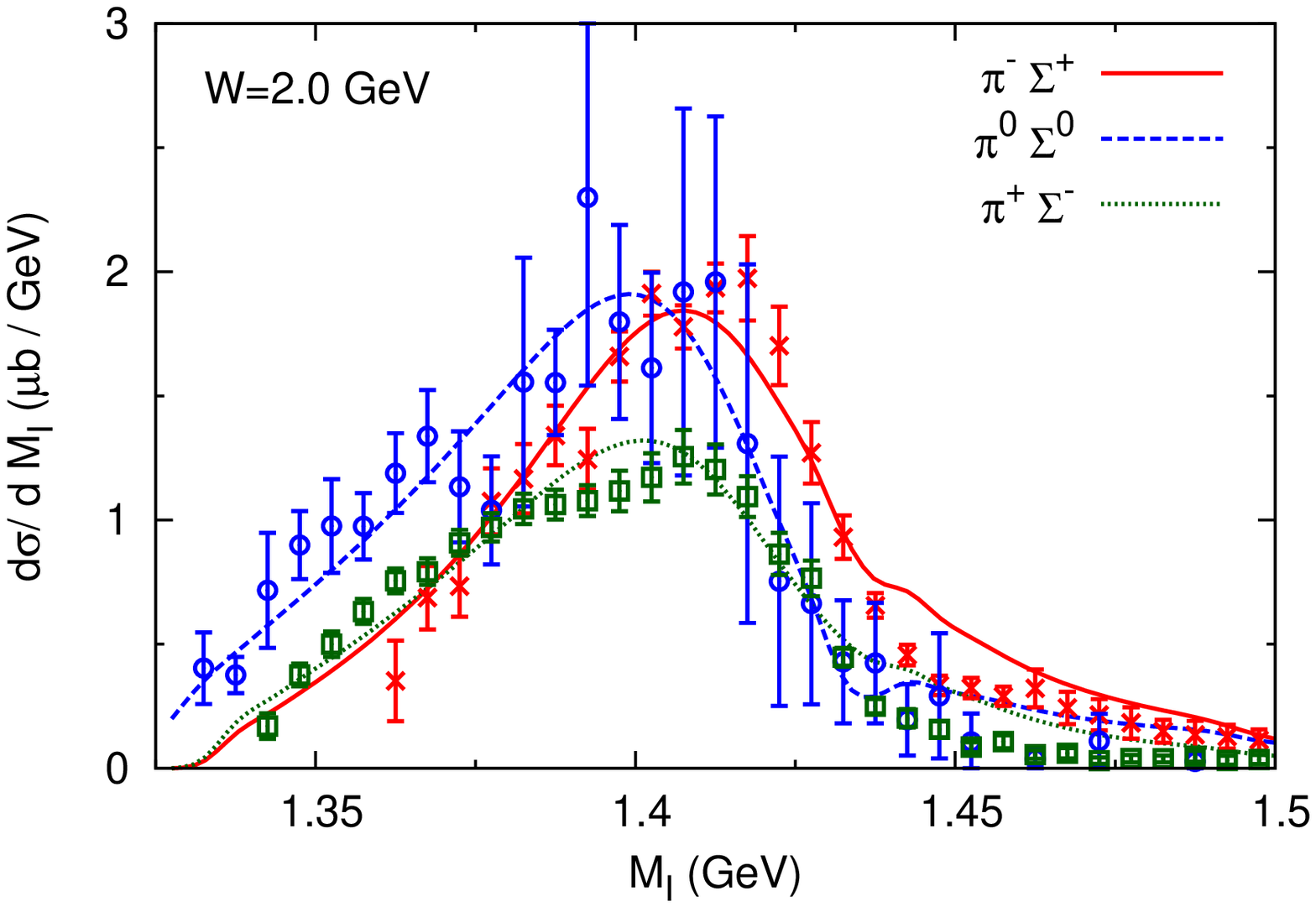}
\includegraphics[clip,width=0.45\textwidth]{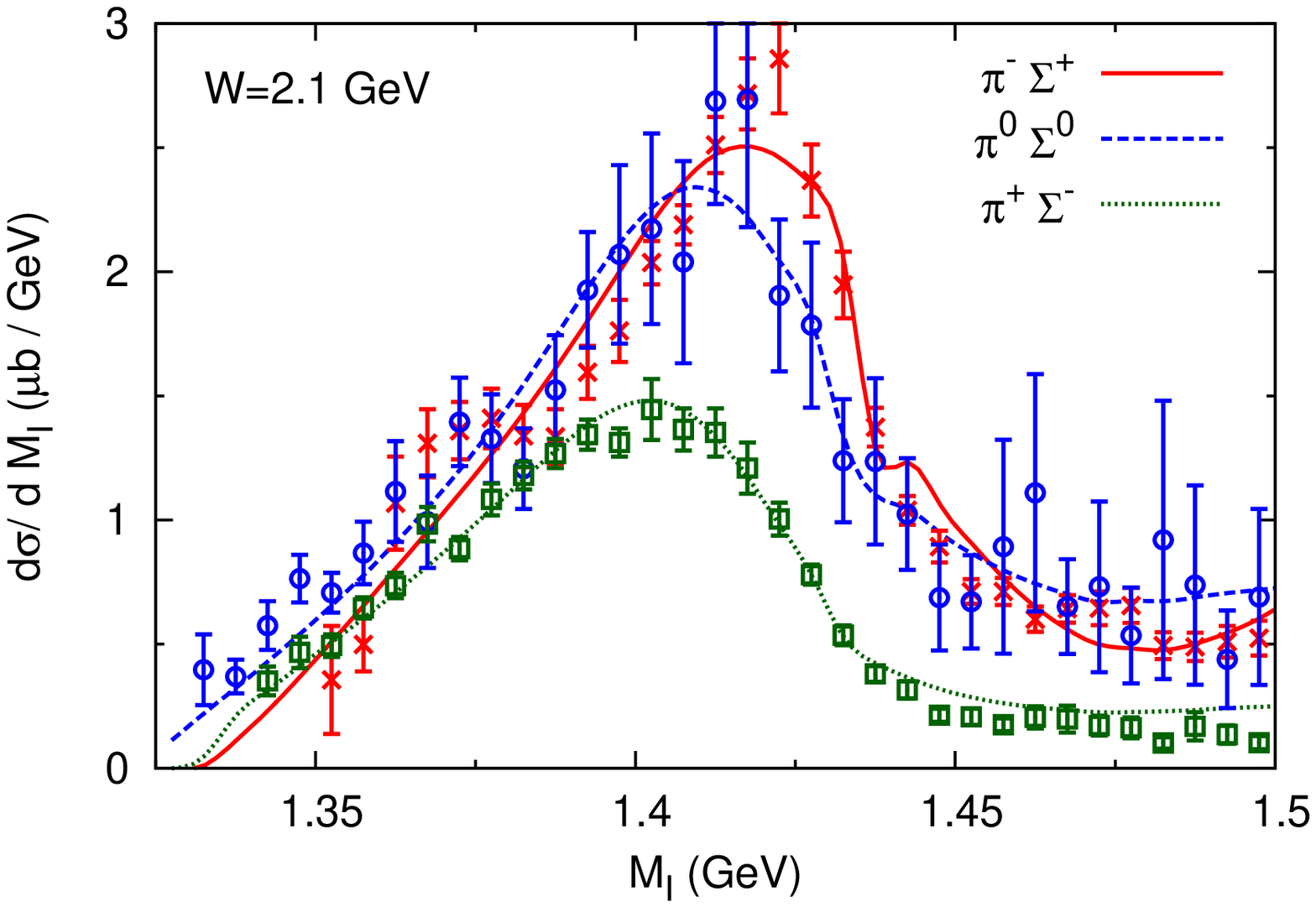}
\includegraphics[clip,width=0.45\textwidth]{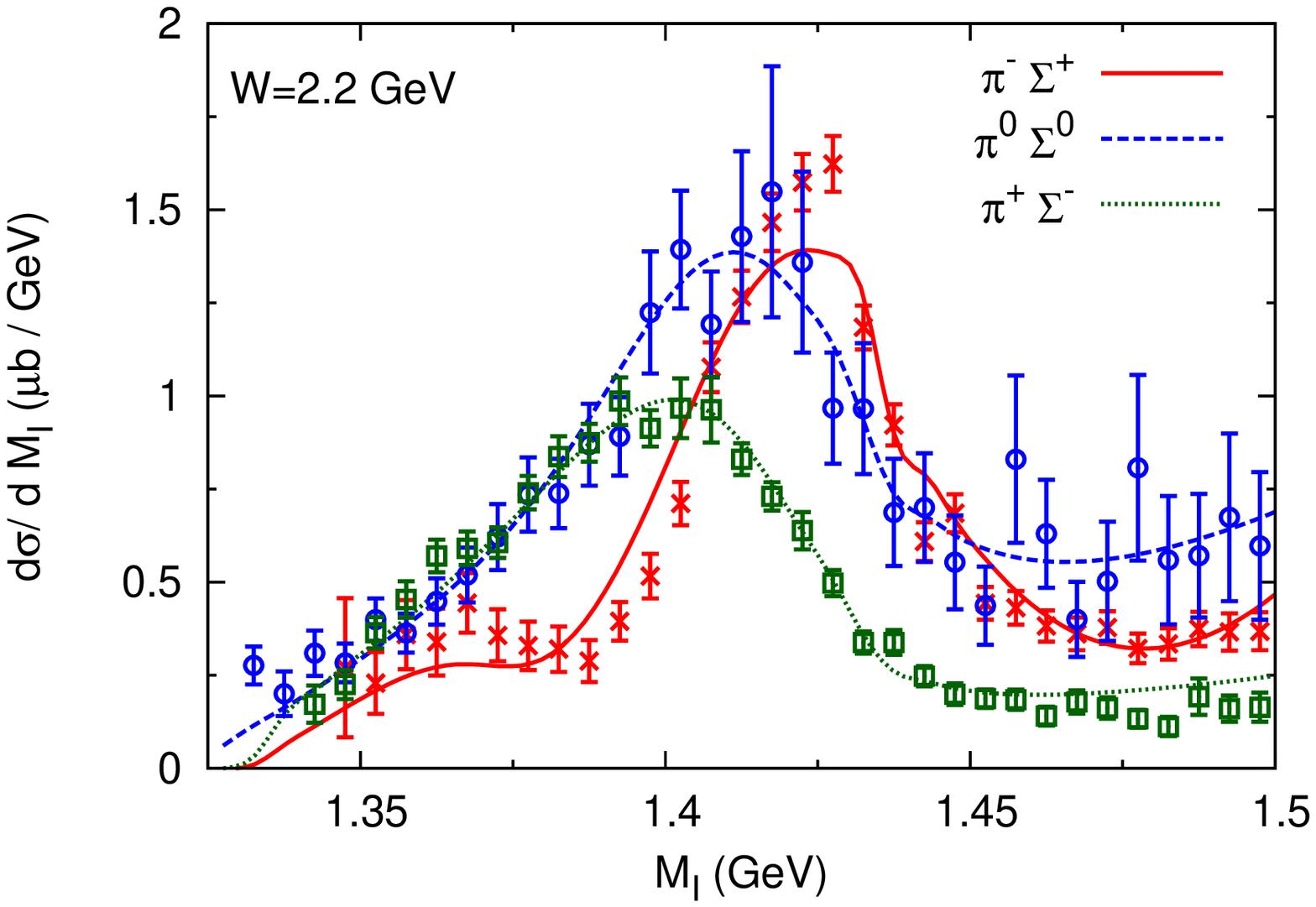}
\includegraphics[clip,width=0.45\textwidth]{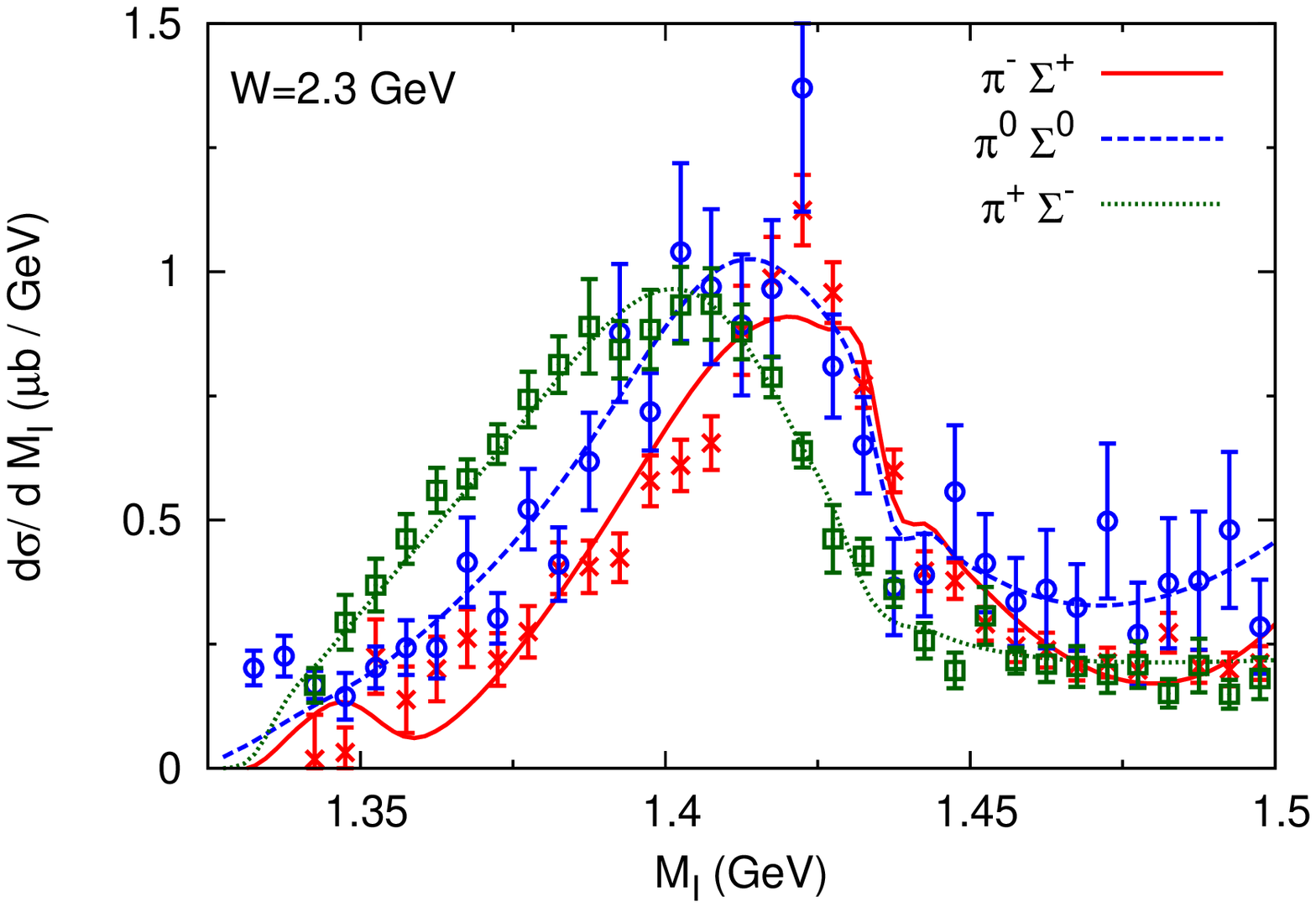}
\caption{\label{fig:bw} (Color online)
Comparison of $\pi\Sigma$ line-shapes from single Breit-Wigner model with
 data~\cite{Moriya:2013eb} at $W=2.0$, 2.1, 2.2 and 2.3~GeV.
See also the caption in Fig.~\ref{fig:2gev}.
}
\end{figure}
Although the quality of the fit is a little worse than the previous model with
the chiral unitary amplitude (Fig.~\ref{fig:2gev}),
still it is an acceptable level. 
Thus, the line-shape data for  $\gamma p\to K^+\pi\Sigma$ only do not
rule out the possibility of single pole solution for $\Lambda(1405)$.
The fit gives $M_{BW}=1412$~MeV and $\Gamma_{BW}=67$~MeV that are close
to the middle of the two poles from the chiral unitary amplitude, but
still closer to the higher mass pole than to the lower one.
As seen in TABLE~\ref{tab:bw-param}, the Breit-Wigner amplitude couples
strongly (weakly) to the $\bar K N$ ($\pi\Sigma$) channel, which is also
similar to the character of the higher mass pole of $\Lambda(1405)$ in
the chiral unitary model.

\subsection{$K^+$ angular distribution}

\begin{figure}
\includegraphics[clip,width=0.45\textwidth]{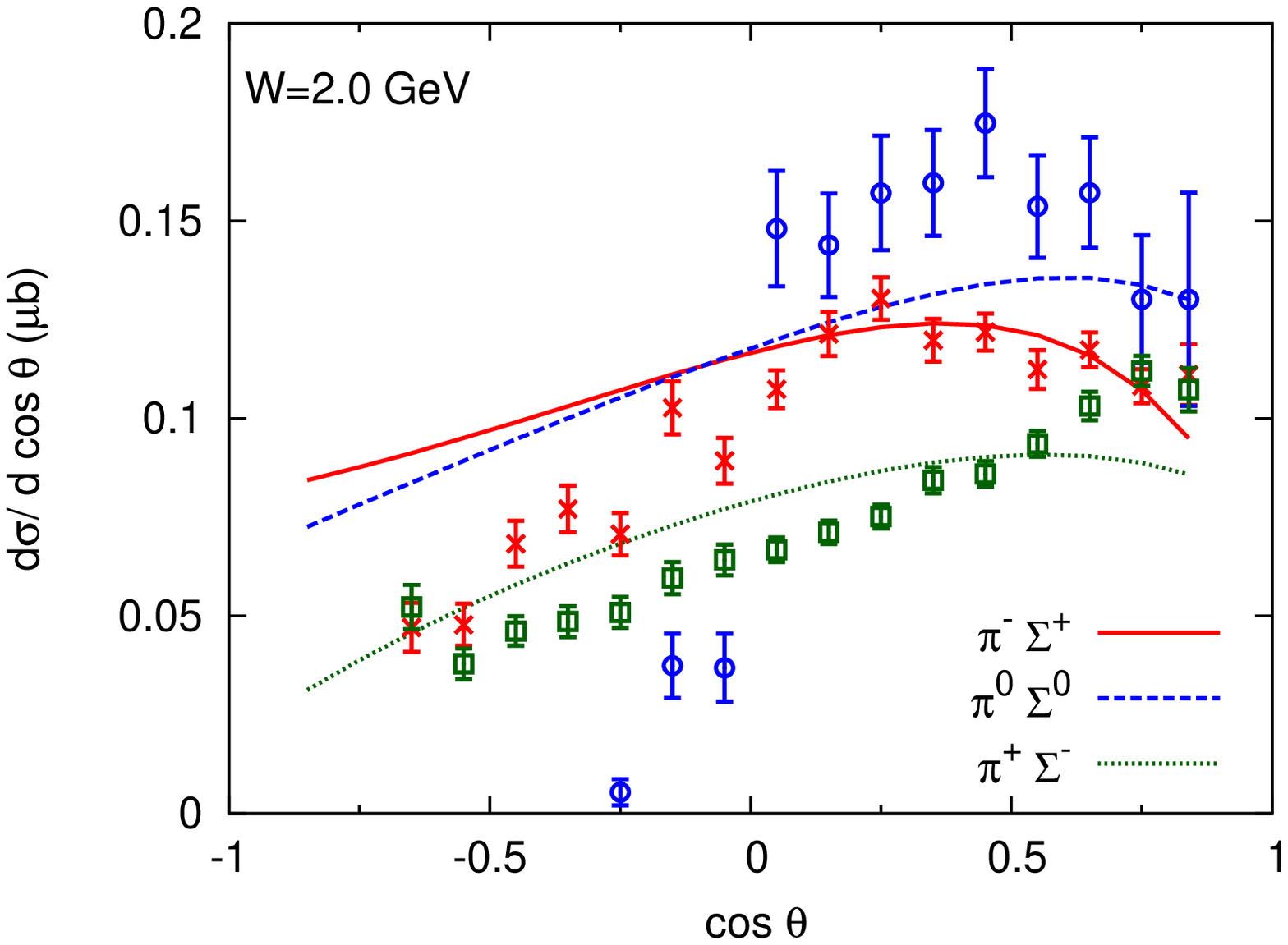}
\includegraphics[clip,width=0.45\textwidth]{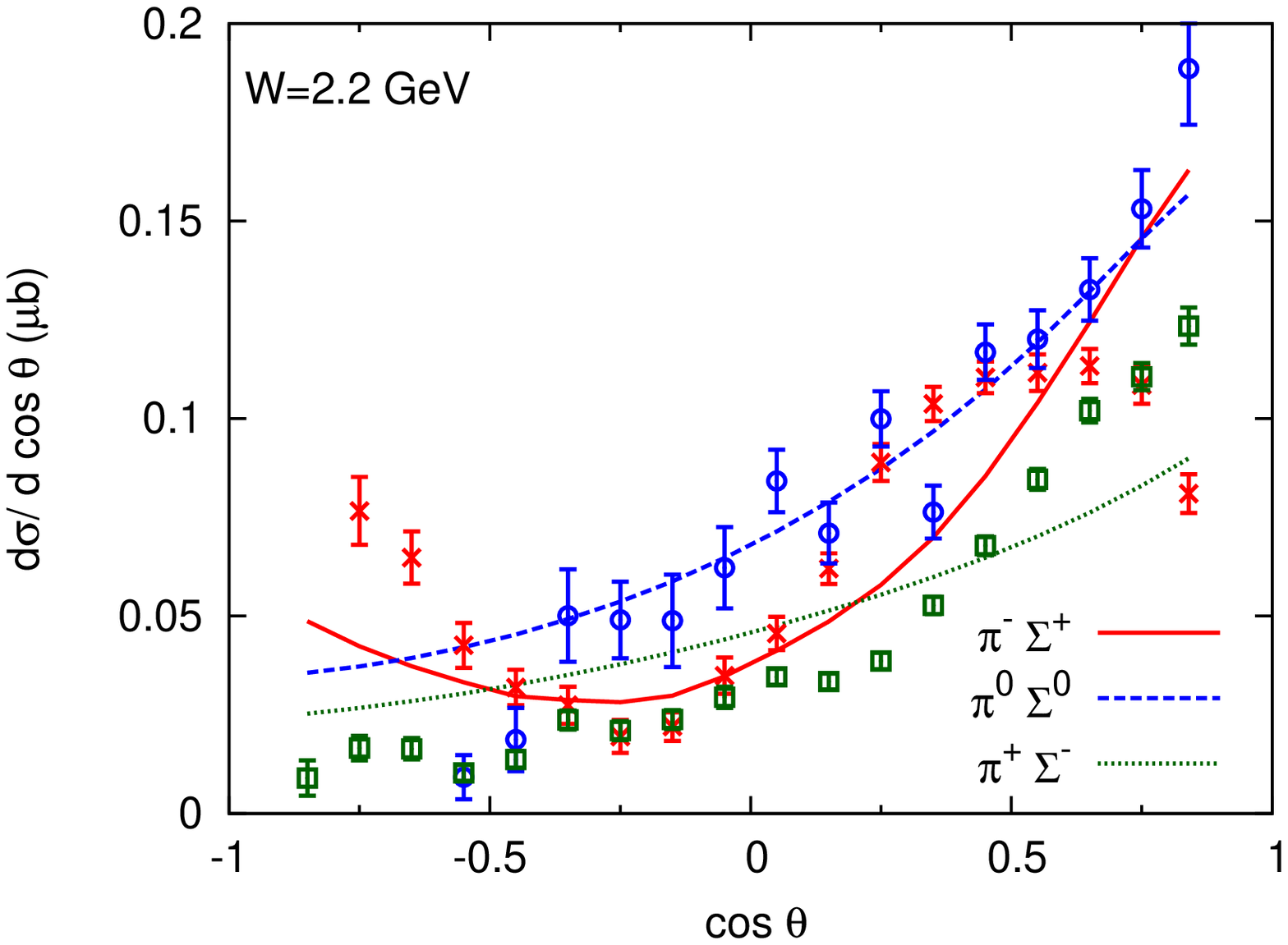}
\caption{\label{fig:angle} (Color online)
Comparison of the $K^+$ angular distributions for $\gamma p\to K^+\pi\Sigma$ 
at $W=2.0, 2.2$~GeV with data from the CLAS~\cite{Moriya:2013hwg}. 
See also the caption in Fig.~\ref{fig:2gev}.
}
\end{figure}

So far, we fitted only the $\pi\Sigma$ line-shape data for the 
$\gamma p\to K^+\pi\Sigma$ reaction.
We found that fitting only the $\pi\Sigma$ line-shape data can lead to
several solutions whose quality of the fit to the line-shape data are
comparable. However, they can have very different $K^+$ angular distribution.
Therefore, $K^+$ angular distribution data will be
useful information to constrain the production mechanism.
Recently the CLAS Collaboration reported data for the $K^+$ angular
distributions~\cite{Moriya:2013hwg}. 
In Figs.~\ref{fig:2gev}-\ref{fig:bg-res}, we actually presented the
$\pi\Sigma$ line-shapes from the model that gives $K^+$ angular distributions
relatively close to the new data of Ref.~\cite{Moriya:2013hwg}. 
Here we use the same model to calculate the $K^+$ angular distributions, and
show them in Fig.~\ref{fig:angle}.
At $W=2.2$~GeV, our model captures overall trend of the data. 
However, for the $\gamma p\to K^+\pi^0\Sigma^0$ reaction at $W=2.0$~GeV, 
there is a sharp rise in the data at $\cos\theta\sim 0$ while rather
smooth behavior is found in the calculated counterpart. 
We actually tried fitting the $K^+$ angular distributions data, but this
sharply rising behavior cannot be fitted with the current setup. 
It seems that we need to search for a mechanism that is responsible for
this behavior. 
We leave such a more detailed analysis of the $K^+$ angular distribution
to a future work.

\section{Summary}
\label{sec:summary}

We calculated the $\pi\Sigma$ line-shapes of the photoproduction 
process $\gamma p \to K^{+} \pi\Sigma$. 
This was motivated by the recent CLAS collaboration's report~\cite{Moriya:2013eb} that found 
peaks due to $\Lambda(1405)$ excitation in the $\pi\Sigma$ line-shapes.
We employed the scattering amplitudes from the chiral unitary model to
describe the final state interaction where the $\Lambda(1405)$ is
excited.
For the tree-level photo-production mechanisms,
we introduced a gauge invariant model that consists of the gauged Weinberg-Tomozawa
terms, the gauged Born terms and the vector-meson exchange terms. 
We also introduced freedom to fit the data such as
contact terms modelling short-range dynamics like baryon resonances, 
and subtraction parameters that can be different from those determined 
in the chiral unitary model. 
These are necessary to reproduce 
the $\pi\Sigma$ line-shape data from the CLAS.
This implies that
the mechanism for the $\Lambda(1405)$ photo-production is not so 
simple that the several important terms interfere. 
It is noted that
we do not adjust any parameters (subtraction constants, couplings) of the chiral
unitary amplitudes in the fit.

Our model reproduces the $\pi\Sigma$ line-shape data quite well.
Breaking down the calculated $\pi\Sigma$ line-shape into 
contributions from each production mechanism,
we found that the contribution from the gauged Weinberg-Tomozawa terms
is not so important due to rather destructive interference between the terms.
More important contributions are from 
the born and vector-meson exchange terms.
In addition, the short range contact terms give large contributions
to reproduce the $\pi\Sigma$ line-shape data.
We also decomposed
the calculated $\pi\Sigma$ line-shapes into the resonant 
and nonresonant parts for each charge state of the $\pi\Sigma$ channels. 
We found that even though the resonant part dominates the spectra and 
generates the peak structure, 
the nonresonant background contribution is not so negligible and its sizable
effect shifts the $\Lambda(1405)$ peak position. The direction of the 
shift is a consequence of complicated interference of many terms and
it is hard to pin down the main mechanism for the shift. 
One can say that it is not the case that the shift is 
solely caused by an interference between the $I=0$ and $I=1$ components,
because one can see the shift of the peak position also in the $\pi^{0}\Sigma^{0}$
channel. 

We also made a check of the $\Lambda(1405)$ amplitude obtained 
by the chiral unitary model.
We refitted the $\pi\Sigma$ line-shape data
with a single Breit-Wigner amplitude for the $\Lambda(1405)$ 
amplitude instead of those from the chiral unitary model.
The quality of the fit is fairly good, indicating that the $\pi\Sigma$
line-shape data alone do not rule out a one-pole solution for $\Lambda(1405)$.
We found that the Breit-Wigner mass and width
obtained by the fit are closer to the higher pole position of the
$\Lambda(1405)$ in the chiral unitary model.
Also the Breit-Wigner amplitude strongly couples to the $\bar K N$
channel, sharing the similar property of the higher mass pole.
This implies that, in the model including the chiral unitary amplitude, 
the higher $\Lambda(1405)$ resonance pole 
plays a more important role for the photoproduction.

In future work, we will simultaneously analyze data for the
$\pi\Sigma$ line-shape and the $K^+$ angular distribution, and then extract
$\Lambda(1405)$ pole(s).
We presented the $K^+$ angular distribution from the current
model obtained by fitting only the line-shape data.
Although our model captures overall behavior of the data in many cases,
there is also a qualitative difference that cannot be fixed by fitting
with the current setup. 
Identifying a mechanism that can fill the difference will be a challenge
in the future work. 
Also an important task is to address a model-dependence of the extracted 
$\Lambda(1405)$ pole(s) because we are using a rather phenomenological
production mechanisms.
This can be done by using contact terms and/or form factors of different
forms.

\begin{acknowledgments}
We thank Reinhard Schumacher and Kei Moriya for useful discussions.
SXN is the Yukawa Fellow and his work is supported in part by Yukawa
Memorial Foundation, by the Yukawa International　Program for Quark-hadron
Sciences (YIPQS), and by Grants-in-Aid for the global COE program “The
Next Generation of Physics, Spun from Universality and Emergence” from
MEXT. This work was partially supported by the Grants-in-Aid for Scientific Research 
(No.\ 25400254 and No.\ 24105706).
\end{acknowledgments}

\appendix

\section{Partial wave expansion}
\label{app:pw}

We summarize a partial wave expansion of an amplitude for 
the $\gamma + p \to K^+ + M + B$ ($M$: meson; $B$: baryon) reaction
in the center-of-mass system of $MB$,
with respect to the relative motion of $MB$.
We denote the amplitude by $T$ and its partial wave
$T_{JL}$ where $J$ and $L$ are total and orbital angular momenta of
$MB$, respectively.
Here we show equations for 
the partial waves relevant to this work; 
$L=0,1$.

The relation between $T_{JL}$ and $T$
for ($J$, $L$)=(1/2,0) is
\begin{eqnarray}
T_{JL} = {1\over 4\pi} \int d\Omega_{\hat{\bm k}_M}\ T \,
\end{eqnarray}
where $\bm k_M$ is the momentum for $M$ 
and $\hat{\bm k}_M ={\bm k}_M /|{\bm k}_M|$.
For ($J$, $L$)=(1/2,1), we have
\begin{eqnarray}
T_{JL} =  {1\over 4\pi}\int d\Omega_{\hat{\bm k}_M}\ \bm\sigma\cdot \hat{\bm k}_M\, T \ ,
\end{eqnarray}
and for ($J$, $L$)=(3/2,1),
\begin{eqnarray}
\bm T_{JL}\cdot \bm n = 
{1\over 4\pi} \int d\Omega_{\hat{\bm k}_M}\
3 \bm S\cdot \hat{\bm n}
\bm S^\dagger\cdot \hat{\bm k}_M \,
T \ ,
\end{eqnarray}
where $\bm n$ is an arbitrary vector.
The operator $S^\dagger$ ($S$) is a baryon spin transition operator from
spin 1/2 to 3/2 (3/2 to 1/2), and it can be expressed by
\begin{eqnarray}
 \bm S\cdot \bm a \bm S^\dagger\cdot \bm b
=  {2\over 3}\bm a \cdot \bm b -{i\over 3}
\bm \sigma \cdot \bm a \times \bm b \ .
\end{eqnarray}
With the partial wave amplitudes $T_{JL}$ defined above, 
the original amplitude $T$ is written as
\begin{eqnarray}
T = 
T_{1/2,  0} +
T_{1/2,  1} \bm\sigma\cdot \hat{\bm k}_M +
\bm T_{3/2, 1}\cdot \hat{\bm k}_M + ({\rm higher\ partial\ waves}) \ .
\end{eqnarray}
A partial wave expansion of production potentials 
[$V^j_\alpha$ in Eq.~(\ref{eq:tamp})] can be done in the same manner,
leading to $V^{j}_{\alpha;JL}$ in Eq.~(\ref{eq:pw}).

\section{ Model Lagrangians }
\label{app:lag}

We present a set of Lagrangians from which we derive photo-production
mechanisms graphically shown in Figs.~\ref{fig:wt}-\ref{fig:vec}.
We follow the convention of Bjorken-Drell.
We use symbols $B$, $M$, $V$ and 
$A$ to denote octet baryon, octet meson, nonet vector meson, and
electromagnetic fields, respectively.
Also, we use curly symbols to denote creation or annihilation operators.
For example, ${\cal B}$ is the annihilation operator contained in $B$, and
its normalization is 
$\bra{0} {\cal B}\ket{B}=1$.

\subsection{Hadronic interactions}

We work with the lowest order chiral Lagrangian for the octet
pseudoscalar mesons ($M$) coupled to the octet $1/2^+$ baryons ($B$),
as given by~\footnote{When $B$, $V$, and $M$ are enclosed by the trace
brackets, they are SU(3) matrix. Otherwise, they are understood to be
one of particles contained in the SU(3) matrix elements.
The same applies to the curly symbols for those fields.
}
\begin{eqnarray}
   {\cal L}_\chi &=& \langle \bar{B} i \gamma^{\mu} \nabla_{\mu} B
    \rangle  - M_B \langle \bar{B} B\rangle  \nonumber \\
    &&  + \frac{1}{2} D \left\langle \bar{B} \gamma_5\gamma^{\mu}  \left\{
     u_{\mu}, B \right\} \right\rangle + \frac{1}{2} F \left\langle \bar{B}
     \gamma_5\gamma^{\mu}  \left[u_{\mu}, B\right] \right\rangle  \ ,
    \label{chiralLag}
\end{eqnarray}
where the symbol $\langle \, \rangle$ denotes the trace of SU(3) flavor
matrices, $M_B$ is the baryon mass and
\begin{eqnarray}
  \nabla_{\mu} B &=& \partial_{\mu} B + [\Gamma_{\mu}, B] \nonumber \ ,\\
  \Gamma_{\mu} &=& \frac{1}{2} (u^\dagger \partial_{\mu} u + u\, \partial_{\mu}
      u^\dagger) \nonumber \ , \\
  U &=& u^2 = {\rm exp} (i \sqrt{2} M / f) \ , \\
  u_{\mu} &=& i u ^\dagger \partial_{\mu} U u^\dagger \nonumber \ . 
\end{eqnarray}
For the couplings $D$ and $F$, we use $D=0.85$, $F=0.52$, and 
$f=1.15 f_\pi$ with $f_\pi=93$~MeV.

The meson and baryon fields in the SU(3) matrix form are
\begin{equation}
M =
\left(
\begin{array}{ccc}
\frac{1}{\sqrt{2}} \pi^0 + \frac{1}{\sqrt{6}} \eta & \pi^+ & K^+ \\
\pi^- & - \frac{1}{\sqrt{2}} \pi^0 + \frac{1}{\sqrt{6}} \eta & K^0 \\
K^- & \bar{K}^0 & - \frac{2}{\sqrt{6}} \eta
\end{array}
\right) \ ,
\end{equation}
\begin{equation}
B =
\left(
\begin{array}{ccc}
\frac{1}{\sqrt{2}} \Sigma^0 + \frac{1}{\sqrt{6}} \Lambda &
\Sigma^+ & p \\
\Sigma^- & - \frac{1}{\sqrt{2}} \Sigma^0 + \frac{1}{\sqrt{6}} \Lambda & n \\
\Xi^- & \Xi^0 & - \frac{2}{\sqrt{6}} \Lambda
\end{array}
\right) \ .
\label{eq:baryon}
\end{equation}
From Eq.~(\ref{chiralLag}),
we will particularly use the $BBMM$ interaction, as contained in the
covariant derivative, given by
\begin{equation}
   {\cal L}_{BBMM} = \left\langle \bar{B} i \gamma^{\mu} \frac{1}{4 f^2}
   [(M\, \partial_{\mu} M - \partial_{\mu} M M) B
   - B (M\, \partial_{\mu} M - \partial_{\mu} M M)] 
   \right\rangle \ , \label{lowest}
\end{equation}
and also use the $BBM$ interaction, as in $D$ and $F$ terms, as
\begin{equation}
   {\cal L}_{BBM} =
      - \sqrt{2} g^{\rm oct}_{PS} \left(\alpha_{PS} \left\langle \bar{B} \gamma_5\gamma^{\mu}  B
     \partial_{\mu}M  \right\rangle_F
+(1-\alpha_{PS}) \left\langle \bar{B}
		  \gamma_5\gamma^{\mu}  B \partial_{\mu}M  \right\rangle_D \right)  \ ,
\label{eq:bbm}
\end{equation}
where $g^{\rm oct}_{PS}=(D+F)/(2f)$ and $\alpha_{PS}=F/(D+F)$, and we have
introduced the traces $\langle\rangle_F$ and $\langle\rangle_D$ defined
by
\begin{eqnarray}
 \left\langle \bar B BM \right\rangle_F &=&  \left\langle \bar B
  [M, B] \right\rangle\\
 \left\langle \bar B BM \right\rangle_D &=&  \left\langle \bar B
  \{M, B\} \right\rangle\ - {2\over 3} \left\langle \bar B B
					  \right\rangle \left\langle M \right\rangle\
.
\end{eqnarray}
We will also use a notation defined by
\begin{eqnarray}
 \left\langle \bar B BM \right\rangle_S &=& 
\left\langle \bar B B  \right\rangle
\left\langle M \right\rangle\ .
\end{eqnarray}
From here, we discuss interactions involving the nonet vector mesons.
In the SU(3) matrix form, the vector meson nonet is given by 
\begin{equation}
V =
\left(
\begin{array}{ccc}
\frac{1}{\sqrt{2}} \rho^0 + \frac{1}{\sqrt{2}} \omega & \rho^+ & K^{*+} \\
\rho^- & - \frac{1}{\sqrt{2}} \rho^0 + \frac{1}{\sqrt{2}} \omega & K^{*0} \\
K^{*-} & \bar{K}^{*0} & \phi
\end{array}
\right) \ ,
\end{equation}
where the ideal mixing between the neutral vector mesons is assumed.
With the matrix, the $VMM$ interactions we use are
\begin{equation}
\label{eq:vmm}
   {\cal L}_{VMM} = - i \sqrt{2} g 
\left\langle V_\mu (\partial^\mu M M - M \partial^\mu M)\right\rangle
 \ ,
\end{equation}
where the coupling $g$ is related to the $\rho\pi\pi$ coupling by
$g_{\rho\pi\pi}=2 g$, and we use $g_{\rho\pi\pi}=6.0$ determined from
the $\rho\to\pi\pi$ decay width.
The vector part of the $BBV$ interactions are given by
\begin{equation}
\label{eq:bbv}
   {\cal L}^v_{BBV} =
      - {\sqrt{2}} g^{\rm oct}_V  \left(\alpha_V\left\langle \bar{B} \gamma^{\mu} B
     V_{\mu} \right\rangle_F + (1-\alpha_V)  \left\langle \bar{B}
     \gamma^{\mu} B V_{\mu}  \right\rangle_D \right) 
- {g^{\rm sin}_V \over \sqrt{3}} \left\langle \bar{B} \gamma^{\mu} B
     V_{\mu} \right\rangle_S
 \ .
\end{equation}
We use the coupling constants
$g^{\rm oct}_V=g=g_{\rho NN}$, 
and $\alpha_V=1$ from the universality assumption.
The relative phase between the $VMM$ and $BBV$ interactions is also
fixed by the universality.
We also use 
$g^{\rm sin}_V=\sqrt{6}\, g^{\rm oct}_V$, so that $g_{\phi NN}=0$
and $g_{\omega NN}=3 g_{\rho NN}$.
The $BBV$ interactions also contain the tensor coupling, as seen in 
the common expressions for the $\rho NN$ and $\omega NN$ interactions:
\begin{eqnarray}
 {\cal L}_{\rho NN} &=&
      - g_{\rho NN}  
 \left( \bar{N} \gamma^{\mu} \bm{\tau} N    \cdot \bm{\rho}_{\mu} 
 - {\kappa_{\rho}\over 2 M_p}
\bar{N} \sigma^{\mu\nu} \bm{\tau} N    \cdot \partial_\nu\bm{\rho}_{\mu}
 \right) \ , \\
 {\cal L}_{\omega NN} &=&
      - g_{\omega NN}  
 \left( \bar{N} \gamma^{\mu} N   \omega_{\mu} 
 - {\kappa_{\omega}\over 2 M_p}
\bar{N} \sigma^{\mu\nu} N    \partial_\nu \omega_{\mu}
 \right) \ ,
\end{eqnarray}
where $M_p$ is the proton mass.
We use $\kappa_\rho=2$ and $\kappa_\omega=0$, based on an average of 
$\pi N$ and $\gamma N$ reaction models~\cite{tensor}.
The tensor couplings for the other vector mesons are fixed using
the SU(3) relation for the magnetic coupling, and an explicit expression
will be given later in Eq.~(\ref{eq:tensor}).
The $BBVM$ interactions are given by
\begin{equation}
\label{eq:bbvm}
   {\cal L}_{BBVM} =
    2 i g^{\rm oct}_V g^{\rm oct}_{PS}  \left(\alpha_{PS}\left\langle \bar{B} \gamma_5\gamma^{\mu} B
     [M,V_{\mu}] \right\rangle_F + (1-\alpha_{PS})  \left\langle \bar{B}
     \gamma_5\gamma^{\mu} B [M,V_{\mu}]  \right\rangle_D \right)  \ .
\end{equation}
The $VVM$ interactions we use are based on the hidden local symmetry
model~\cite{HLS}, and are given by
\begin{equation}
\label{eq:vvm}
   {\cal L}_{VVM} = g^2 C\epsilon^{\alpha\beta\gamma\delta}
\left\langle \partial_\alpha V_\beta \partial_\gamma V_\delta M
\right\rangle
 \ ,
\end{equation}
where $C=-3/(2\sqrt{2}\pi^2f)$, and we use the convention, $\epsilon^{0123}=+1$.

\subsection{Electromagnetic interactions}

The photon coupling to the baryonic current is given by
\begin{eqnarray}
L_{\gamma BB'} &=& 
- e \bar B \left[ 
Q_{BB'} \sla{A} - \frac{{\kappa}_{BB'}}{2M_p}\sigma^{\mu\nu} (\partial_\nu A_\mu)
\right]B' \, .
\end{eqnarray}
The symbol $Q_{BB'}$ is the electric charge (in unit of $e=|e|$) of a baryon $B$ for $B=B'$, 
but zero otherwise.
The anomalous magnetic moment is denoted by ${\kappa}_{BB'}$ for
which we use experimental values listed in the Particle Data
Group~\cite{pdg}~\footnote{
The magnetic moment for $\Sigma^0$ has not been measured, and we use a
quark model prediction~\cite{pdg}.
We also use the quark model to fix
the sign for $\kappa_{\Lambda\Sigma^0}$.
}.

The photon coupling to the pseudoscalar meson current is given by
\begin{eqnarray}
L_{\gamma \pi\pi} &=& 
-i e \left[ \pi^-\partial^\mu \pi^+ - (\partial^\mu \pi^-)\pi^+ \right]
A_\mu \, ,
\\
L_{\gamma KK} &=& 
-i e \left[ K^-\partial^\mu K^+ - (\partial^\mu K^-)K^+ \right] A_\mu \,.
\end{eqnarray}

The minimal substitutions ($\partial_{\mu} \to ie A_\mu [Q,]$)
to the $BBMM$~[Eq.~(\ref{lowest})] and $BBM$~[Eq.~(\ref{eq:bbm})] interactions
respectively give
\begin{equation}
   {\cal L}_{\gamma BBMM} = - \frac{e}{4 f^2} \left\langle \bar{B} \gamma^{\mu} 
  \left\{ (M\,  [Q,M] -   [Q,M] M) B
   - B (M\, [Q,M] -  [Q,M] M) \right\}
   \right\rangle A_{\mu} \ , \label{lowest-em}
\end{equation}
and
\begin{equation}
   {\cal L}_{\gamma BBM} =
      - ie \sqrt{2} g^{\rm oct}_{PS} \left(\alpha_{PS} \left\langle \bar{B} \gamma_5\gamma^{\mu}  B
      [Q,M]  \right\rangle_F
+(1-\alpha_{PS}) \left\langle \bar{B}
      \gamma_5\gamma^{\mu} B [Q,M]  \right\rangle_D \right) A_{\mu} \ ,
\end{equation}
where $Q$ is the quark charge matrix $Q={\rm diag} (2/3,-1/3,-1/3)$.

The electromagnetic interactions involving the vector mesons are due to
the U(1) axial anomaly, and are given by
\begin{eqnarray}
L_{\gamma VM} &=& g_{\gamma VM} \epsilon^{\alpha\beta\gamma\delta}
 M \partial_\alpha A_\beta \partial_\gamma V_\delta \ .
\label{eq:gvp}
\end{eqnarray}
The couplings $g_{\gamma VM}$ are determined by experimental 
$V\to \gamma M$ decay widths~\cite{pdg}, and the relative phases are fixed by
the SU(3) relation.
The numerical values for $g_{\gamma VM}$ are given in
TABLE~\ref{tab:gvp}.~\footnote{
Although we find $|g_{\gamma \phi\pi^0}|=0.04$ GeV$^{-1}$ from data, 
the SU(3) predicts it to be zero, and we cannot find its phase. 
In this work, we set it to be zero.
}

\begin{table}
\caption{\label{tab:gvp} Coupling constants for $L_{\gamma VM}$ given in Eq.~(\ref{eq:gvp}).
}
\renewcommand{\arraystretch}{1.2}
\tabcolsep=1.4mm
\begin{tabular}{c|cccccccccc}\hline
$\gamma VM$&
${\gamma K^{*\pm}K^{\mp}}$&
${\gamma K^{*0}\bar{K}^{0}}$&
${\gamma \bar{K}^{*0}{K}^{0}}$&
${\gamma \omega \eta}$&
${\gamma \omega\pi^0}$&
${\gamma \rho^0 \eta}$&
${\gamma \rho^0\pi^0}$&
${\gamma \rho^\pm \pi^\mp}$&
${\gamma \phi\eta}$&
${\gamma \phi\pi^0}$\\\hline
$g_{\gamma VM}$ (GeV$^{-1}$)
&0.254 &$-0.388$&$-0.388$&0.169&0.736&0.565&0.234&0.221&0.216&0\\\hline
\end{tabular}
\end{table}

\subsection{Matrix elements and coupling constants }
\label{app:coup}

In this subsection, 
we evaluate matrix elements of the Lagrangians defined above,
in order to introduce a coupling constant for a given set of
incoming and outgoing particles. 
The coupling constants will be used to write down the photoproduction
amplitudes in the next section. 
We will often use an index $i\, ({\rm or}\ j)=1, ..., 10$ to specify a pair of
meson and baryon,
$M_iB_i=K^-p, \bar K^0 n, \pi^0\Lambda, \pi^0\Sigma^0, \eta\Lambda, \eta\Sigma^0, \pi^+\Sigma^-, \pi^-\Sigma^+, K^+\Xi^-, K^0\Xi^0$, respectively.
Also we denote four-momenta for $M_i$ and $B_i$ as $k_i$ and $p_i$, respectively.

The matrix element of the $BBMM$ interaction defined in 
Eq.~(\ref{lowest}) is given by
\begin{equation}
\bra{B_j(p_j)M_j(k_j)}  {\cal L}_{BBMM} \ket{B_i(p_i)M_i(k_i)}=
\sqrt{M_{B_i} M_{B_j}\over 4 E_{B_i} E_{B_j} E_{M_i} E_{M_j}}
{C_{ji}\over 4f^2} 
 \bar u_{B_j}(p_j)(\slas{k}_j+\slas{k}_i)u_{B_i}(p_i)
 \ ,
\end{equation}
where the energy for a particle $x$ is denoted by 
$E_x(p_x)=\sqrt{\bm{p}_x^2+M_x^2}$ where $M_x$ is the mass of $x$;
the values of $M_x$ are from Ref.~\cite{pdg}.
The couplings $C_{ji}$ are tabulated in TABLE~1 of Ref.~\cite{OR}.
The matrix element of the $BBM$ interaction defined in 
Eq.~(\ref{eq:bbm}) is given with a coupling constant $D^j_{B}$ as
\begin{equation}
\bra{B_j(p_j)M_j(k_j)} {\cal L}_{BBM} \ket{B(p)}=
i 
\sqrt{M_{B} M_{B_j}\over 2 E_{B} E_{B_j} E_{M_j}}
D^{M_jB_j}_{B}
\bar u_{B_j}(p_j) \gamma_5\slas{k}_j u_{B}(p)
 \ ,
\end{equation}
with
\begin{equation}
D^{M_jB_j}_{B} =
D^j_{B} = 
     -  \sqrt{2} g^{\rm oct}_{PS}\, 
\bra{B_jM_j} 
 \left(\alpha_{PS} \left\langle
						     \bar{\cal B} {\cal B} {\cal M} 
      \right\rangle_F
+(1-\alpha_{PS}) \left\langle \bar{\cal B} {\cal B} {\cal M} 
       \right\rangle_D \right)  
\ket{B}
 \ .
\label{eq:mbb-coup}
\end{equation}
Similarly,
the matrix elements of the $VMM$, $BBV$,  $BBVM$, and $VVM$
interactions defined respectively in Eqs.~(\ref{eq:vmm}), (\ref{eq:bbv}),
(\ref{eq:bbvm}) and (\ref{eq:vvm}) are given below:
\begin{eqnarray}
\bra{M(k)M'(k')} {\cal L}_{VMM} \ket{V(k_V)} &=&
-g_{VMM'} (k^\mu-k'^\mu)\epsilon_\mu^V 
{1\over \sqrt{8 E_{V} E_{M} E_{M'}}}
 \ , 
\end{eqnarray}
where $\epsilon_\mu^V$ is the polarization vector for the vector meson
$V$, and
\begin{equation}
\bra{B'(p')} {\cal L}_{BBV} \ket{B(p)V(k_V)} =
- \sqrt{M_{B} M_{B'}\over 2 E_{B} E_{B'} E_{V}}
 g_{VBB'} \bar u_{B'}(p') 
\left(\gamma_\mu + i{\kappa_{VBB'}\over 2 M_B}\sigma_{\mu\nu}k_V^\nu
\right)
\epsilon^\mu_V 
 u_{B}(p)
 \ , \\
\end{equation}
\begin{eqnarray}
&&\bra{B_j(p_j)M_j(k_j)} {\cal L}_{BBVM} \ket{B(p)V(k_V)} =
i \sqrt{M_{B} M_{B_j}\over 4 E_{B} E_{B_j} E_{V} E_{M_j}}
g^{M_jB_j}_{VB} \bar u_{B_j}(p_j) \slas{\epsilon_V}
\gamma_5
 u_{B}(p)
 \ , \\
&&\bra{V'(k_{V'})M(k)} {\cal L}_{VVM} \ket{V(k_V)} =
g_{VV'M}
\epsilon_{\alpha\beta\gamma\delta}k_{V'}^\alpha\epsilon_{V'}^\beta
k_V^\gamma\epsilon_V^\delta
{1\over \sqrt{8 E_{V} E_{V'} E_{M}}}
 \ ,
\end{eqnarray}
where we have introduced the coupling constants, 
$g_{VMM'}$, $g_{VBB'}$, $\kappa_{VBB'}$, $g^{M_jB_j}_{VB}\, (=g^j_{VB})$
and $g_{VV'M}$,
and they are related to the parameters in the original Lagrangians
through equations similar to Eq.~(\ref{eq:mbb-coup}).

Finally, the tensor couplings for the vector mesons ($\kappa_{VBB'}$) are fixed using
the SU(3) relation for the magnetic coupling
($G^m_{VBB'}=g_{VBB'}(1+\kappa_{VBB'})$), as given by
\begin{equation}
\label{eq:tensor}
G^m_{B_1B_2V_1} =
\bra{B_2}\left[      - {\sqrt{2}} g^{\rm oct}_m
	  \left(\alpha^m_V\left\langle \bar{\cal B}{\cal B}{\cal V}
      \right\rangle_F + (1-\alpha^m_V)  \left\langle \bar{\cal B}{\cal B}{\cal V} \right\rangle_D \right) 
- {g^{\rm sin}_m \over \sqrt{3}} \left\langle \bar{\cal B}{\cal B}{\cal V}
				 \right\rangle_S \right] \ket{B_1V_1}
 \ ,
\end{equation}
where we use $g^{\rm oct}_m=-9$, $g^{\rm sin}_m=-8.82$, and the static SU(6) value, $\alpha^m_V=0.4$.

\section{Tree-level photoproduction amplitudes}
\label{app:ph-amp}

We present expressions for our tree-level photo-production amplitudes
for $\gamma (q) + p (p) \to K^+(k) + M_j (k_j) + B_j (p')$
where the four-momentum for each particle is given in the parentheses.
The expressions are gauge invariant for
${\cal O} (1)$ and
${\cal O} (1/M_B)$ terms.
The final state is specified by the index $j$
introduced in the previous section.
We denote each amplitude by $V^j_x$ where the label
$x=\ref{fig:wt}({\rm a}), \ref{fig:wt}({\rm b}), ... $ 
specifies the mechanism by referring to the diagram with the same label
in Figs.~\ref{fig:wt}-\ref{fig:vec}.
In what follows, kinematical variables (momentum, energy, polarization
vector) are understood be those in the $M_jB_j$ center-of-mass system,
and omit $*$ that has been used in Eqs.~(\ref{eq:xs}) and (\ref{eq:ls})
for simplicity.
All coupling constants appearing in the expressions have been defined in
Appendix~\ref{app:lag}.

First we introduce building blocks, $F_1, ..., F_4$, to express $V^j_x$ 
in a concise manner:
\begin{eqnarray}
F_1 &=&
{M_p\over E_p(p+q)}{1\over q+E_p(p)-E_p(p+q)}
\left( {\bm{p}+(\bm{p}+\bm{q})\over 2 M_p} + i
 {(1+\kappa_p)(\bm{\sigma}\times\bm{q})\over 2 M_p}\right)\cdot
\bm{\epsilon}\ , \\
F_2 &=&
\left( Q_{B_j}{(\bm{p}'-\bm q)+\bm{p}'\over 2 M_{B_j}} + i
 {(Q_{B_j}+\kappa_{B_j})(\bm{\sigma}\times\bm{q})\over 2 M_p}\right)\cdot \bm{\epsilon}
\nonumber \\
&&\times {M_{B_j}\over E_{B_j}(p'-q)}{1\over E_p(p)-E_{K^+}(k)-E_{M_j}(k_j)-E_{B_j}(p'-q)} 
\ , \\
\label{eq:f3}
F_3 &=&
{ (\bm k - (\bm q - \bm k))\cdot \bm{\epsilon}  \over (q-k)^2 - M_{K^+}^2} 
\ , \\
\label{eq:f4}
F_4 &=&
{ (\bm k_j - (\bm q - \bm k_j))\cdot \bm{\epsilon}  \over (q-k_j)^2 - M_{M_j}^2} \ ,
\end{eqnarray}
where the electric charge of a particle $x$ in unit of $e$ is denoted by
$Q_x$.
The polarization vector of the photon is denoted by $\bm \epsilon$.
The squared momenta in the denominators of $F_3$ and $F_4$
are Lorentz scalar.

\subsection{Gauged Weinberg-Tomozawa terms}

With the minimal substitution to the Weinberg-Tomozawa terms, the
resulting photo-production amplitudes shown in Fig.~\ref{fig:wt} are
given by the following expressions:

\begin{eqnarray}
\label{eq:wt-a}
V^j_{\ref{fig:wt}\rm (a)}  &=& e {C_{j1}\over 4f^2} 
\left[ (k_j^0-k^0)
+{1\over 2 M_B}\left\{ (\bm{k}-\bm{k}_j)^2+ 2i
		\bm{k}\times\bm{k}_j\cdot\bm\sigma \right\}
\right] 
F_1
\nonumber \\
&& - e {C_{j1}\over 4f^2} {1\over 2 M_B} 
\left(\bm{k}_j-\bm{k} + i\bm\sigma\times(\bm{k}_j-\bm{k})
\right) \cdot \bm{\epsilon} \ , \\
\label{eq:wt-b}
V^j_{\ref{fig:wt}\rm (b)}  &=& 
e 
{C_{j1}\over 4f^2} 
F_2
\left[ (k_j^0-k^0)
+{1\over 2 M_B} \left\{ (\bm k - \bm k_j)\cdot (\bm k - \bm k_j - 2 \bm q)
+ 2i \bm{k}\times\bm{k}_j\cdot\bm\sigma \right\}
\right] \nonumber   \\
&& - e\, Q_{B_j} {C_{j1}\over 4f^2} {1\over 2 M_B} 
\left(\bm{k}_j-\bm{k} - i\bm\sigma\times(\bm{k}_j-\bm{k})
\right) \cdot \bm{\epsilon} \ , \\
\label{eq:wt-c}
V^j_{\ref{fig:wt}\rm (c)}  &=& e {C_{j1}\over 4f^2} 
\left[ k_j^0+k^{'0} 
+{1\over 2 M_B} \left\{ (\bm k' + \bm k_j)^2 
+ 2i (\bm k_j \times \bm k')\cdot \bm\sigma \right\}
\right]
F_3
\ , \\
\label{eq:wt-d}
V^j_{\ref{fig:wt}\rm (d)}  &=&  e Q_{M_j} {C_{j1}\over 4f^2} 
\left[ k^0+k_j^{'0} 
-{1\over 2 M_B} \left\{ (\bm k - \bm k_j - \bm q)\cdot (\bm k + \bm k'_j)
+ 2i (\bm k'_j \times \bm k)\cdot \bm\sigma \right\}
\right]
F_4
\ , \\
V^j_{\ref{fig:wt}\rm (e)}  &=& - e {C_{j1}\over 4f^2} (Q_{M_1}+Q_{M_j}) {1\over 2 M_B}
\left( \bm k - \bm q - \bm k_j + i \bm\sigma\times (\bm q-\bm k-\bm k_j)
\right)\cdot \bm{\epsilon} \ .
\end{eqnarray}
The baryon mass $M_B$ in ${\cal O} (1/M_B)$ terms is set to $M_B=M_p$ in actual numerical calculation.
Momenta for intermediate mesons in Eqs.~(\ref{eq:wt-c}) and (\ref{eq:wt-d})  are
$k'=q-k$, 
$k'_j=q-k_j$, respectively.
The last terms in Eqs.~(\ref{eq:wt-a}) and (\ref{eq:wt-b}) are, in
the time-ordered perturbation, due to the propagation of the anti-baryons.
For channels where $B_j=\Lambda$ or $\Sigma^0$, 
the mechanism of Fig.~\ref{fig:wt} (b) contains 
the $\Lambda$-$\Sigma^0$ mixing mechanism, and the following term needs to be added to
Eq.~(\ref{eq:wt-b}):
\begin{eqnarray}
V^j_{\ref{fig:wt}\rm (b)}  &=& 
e \left( i {\kappa_{\Lambda\Sigma^0} (\bm{\sigma}\times\bm{q})\over 2 M_p}\right)\cdot \bm{\epsilon}
{M_{B_{j'}}\over E_{B_{j'}}(p'-q)}{1\over E_p(p)-E_{K^+}(k)-E_{M_j}(k_j)-E_{B_{j'}}(p'-q)} \nonumber \\
&&\times 
{C_{j'1}\over 4f^2} 
\left[ (k_j^0-k^0)
+{1\over 2 M_B} \left\{ (\bm k - \bm k_j)\cdot (\bm k - \bm k_j - 2 \bm q)
+ 2i \bm{k}\times\bm{k}_j\cdot\bm\sigma \right\}
\right]  \ ,
\end{eqnarray}
where we have used a channel index $j'$ to indicate
$B_{j'}=\Lambda (\Sigma^0)$ for $B_j=\Sigma^0 (\Lambda)$
and $M_{j'}=M_{j}$.

\subsection{Gauged Born terms}

For convenience, 
we introduce some building blocks as follows:
\begin{eqnarray}
O_1 &=& \bm \sigma\cdot\bm k_j\left(1+ {k_j^0\over 2 M_B}\right) \ , \\
O_2 &=& \bm \sigma\cdot\bm k \left(1- {k^0\over 2 M_B}\right) 
+ {k^0\over M_B} \bm \sigma\cdot\bm q \ , \\
O_3 &=& \bm \sigma\cdot\bm k
\left(1-{k^0\over 2M_B}\right) + {k^0\over M_B}\bm \sigma\cdot\bm k_j\ , \\
O_4 &=& \bm \sigma\cdot\bm k_j \left(1+ {k_j^0\over 2 M_B}\right) 
- {k_j^0\over M_B} \bm \sigma\cdot(\bm k - \bm q)\ , \\
S_1^{Y^0} &=& {1 \over q+E_p(p)-E_{K^+}(k)-\tilde M_{Y^0}}\ , \\
S_2^{Y^0} &=& 
{M_{Y^0}\over E_{Y^0}(p-k)}{1\over E_p(p)-E_{K^+}(k)-E_{Y^0}(p-k)}\ , \\
S_3^{B_x} &=& 
{M_{B_x}\over E_{B_x}(p+q-k_j)}{1 \over q+E_p(p)-E_{M_j}(k_j)-E_{B_x}(p+q-k_j)}\ , \\
S_4^{B_x} &=& 
{M_{B_x}\over E_{B_x}(p-k_j)}{1 \over E_p(p)-E_{M_j}(k_j)-E_{B_x}(p-k_j)}\ ,
\end{eqnarray}
where $Y^0$ is either $\Lambda$ or $\Sigma^0$.
The quantity $\tilde M_{Y^0}$ is a 'bare' mass, and we use it only in $S_1^{Y^0}$.
The $p$-wave rescattering following $S_1^{Y^0}$ renormalize the bare
mass to give the physical mass.
Because we use the $p$-wave scattering amplitude from Ref.~\cite{JOR},
we also use the bare mass from Ref.~\cite{JOR}; 
$\tilde M_{\Lambda}=1078$~MeV and $\tilde M_{\Sigma^0}=1104$~MeV.
With the minimal substitution to the Born terms, the
resulting photo-production amplitudes shown in Fig.~\ref{fig:br1} are
given by the following expressions:
\begin{eqnarray}
V^j_{\ref{fig:br1}\rm (a)}  &=& e\, 
\sum_{Y^0} 
D^j_{Y^0} D^1_{Y^0} S_1^{Y^0}
\left[
\left(1- {k^0\over 2 M_B}\right)
O_1
\bm \sigma\cdot\bm k
F_1
+ {k^0\over 2 M_B}
\bm \sigma\cdot\bm k_j 
\bm \sigma\cdot\bm \epsilon
\right]
\ , \\
\label{eq:br-b}
V^j_{\ref{fig:br1}\rm (b)}  &=& e\, 
 \sum_{Y^0} 
D^j_{Y^0} D^1_{Y^0} S_1^{Y^0}
O_1
\bm \sigma\cdot\bm \epsilon \ ,\\
\label{eq:br-c}
V^j_{\ref{fig:br1}\rm (c)}  &=& -e\, 
 \sum_{Y^0} 
D^j_{Y^0} D^1_{Y^0} S_1^{Y^0}
\left(1+ {k'^0\over 2 M_B}\right)
O_1
\bm \sigma\cdot\bm k' 
F_3
\ ,\\
V^j_{\ref{fig:br1}\rm (d)}  &=& 
e\,
 \sum_{Y^0} 
D^j_{Y^0} S_1^{Y^0}
\left[
D^1_{Y^0}
\kappa_{Y^0}\,
S_2^{Y^0}
+ 
 \sum_{Y^{0'}\neq Y^0} 
D^1_{Y^{0'}} 
 \kappa_{\Lambda\Sigma^0}\,
S_2^{Y^{0'}}
\right]
O_1
i { \bm{\sigma}\times\bm{q}\cdot \bm{\epsilon}\over 2
M_p}
O_2
\ ,\\
V^j_{\ref{fig:br1}\rm (e)}  &=& e\, Q_{M_j} 
\sum_{Y^0} 
D^j_{Y^0} D^1_{Y^0} S_2^{Y^0}
\bm \sigma\cdot\bm \epsilon
O_2
\ ,\\
V^j_{\ref{fig:br1}\rm (f)}  &=& -e\, Q_{M_j} 
 \sum_{Y^0} 
D^j_{Y^0} D^1_{Y^0} S_2^{Y^0}
F_4
\left\{\bm \sigma\cdot\bm k'_j + {k_j'^0\over 2 M_B} \bm \sigma\cdot
 (\bm k_j + \bm q)\right\}
O_2
\ ,\\
V^j_{\ref{fig:br1}\rm (g)}  &=& e\, 
 \sum_{Y^0} 
D^j_{Y^0} D^1_{Y^0} S_2^{Y^0}
F_2 
\left\{\left(1 + {k_j^0\over 2 M_B} \right) \bm \sigma\cdot\bm k_j 
+ {k_j^0\over M_B} \bm \sigma\cdot\bm q\right\}
O_2
\nonumber \\
&& - e\, Q_{B_j} 
\sum_{Y^0} 
 D^j_{Y^0} D^1_{Y^0} S_2^{Y^0}
 {k^0_j\over 2 M_B} 
\bm \sigma\cdot\bm \epsilon
\bm \sigma\cdot\bm k
\ ,\\
V^j_{\ref{fig:br1}\rm (h)}  &=& e\, 
\sum_{B_x} 
D^{K^+B_j}_{B_x} D^{M_j p}_{B_x} S_3^{B_x}
O_3
\left\{\left(1+{k_j^0\over 2M_B}\right) 
\bm \sigma\cdot\bm k_j
- {k_j^0\over M_B}\bm \sigma\cdot\bm k\right\}
F_1
\nonumber \\
&& - e\, 
\sum_{B_x} 
D^{K^+B_j}_{B_x} D^{M_j p}_{B_x} S_3^{B_x}
{k_j^0\over 2 M_B}
\bm \sigma\cdot\bm k 
\bm \sigma\cdot\bm \epsilon
\ , \\
V^j_{\ref{fig:br1}\rm (i)}  &=& e\, Q_{M_j}
\sum_{B_x} 
D^{K^+B_j}_{B_x} D^{M_j p}_{B_x} S_3^{B_x}
O_3
 \bm \sigma\cdot\bm \epsilon \ ,\\
V^j_{\ref{fig:br1}\rm (j)}  &=& - e\, Q_{M_j}
 \sum_{B_x} 
D^{K^+B_j}_{B_x} D^{M_j p}_{B_x} S_3^{B_x}
O_3
\left\{\left(1-{k'^0_j\over 2M_B}\right)\bm \sigma\cdot\bm k'_j 
-{k'^0_j\over M_B}\bm \sigma\cdot(\bm k-\bm q)\right\}
F_4
\ ,\\
\label{eq:brn-k}
V^j_{\ref{fig:br1}\rm (k)}  &=& e\,
 \sum_{B_x} 
D^{K^+B_j}_{B_x} D^{M_j p}_{B_x} S_3^{B_x}
O_3
 \left( Q_{B_x}{(\bm{k}-\bm k_j-\bm q)+(\bm k-\bm k_j)\over 2 M_{B_x}} + 
i {(Q_{B_x}+\kappa_{B_x})\,\bm{\sigma}\times\bm{q}\over 2
M_p}\right)\cdot \bm{\epsilon}
S_4^{B_x}
O_4
\nonumber\\
&& - e\, 
 \sum_{B_x} 
Q_{B_x} D^{K^+B_j}_{B_x} D^{M_j p}_{B_x} 
\left[
S_3^{B_x}
{k_j^0\over 2 M_B} 
\bm \sigma\cdot\bm k 
\bm \sigma\cdot\bm \epsilon 
+ S_4^{B_x}
{k^0\over 2 M_B}
 \bm \sigma\cdot\bm \epsilon
\bm \sigma\cdot\bm k_j \right]
\ ,\\
\label{eq:br-l}
V^j_{\ref{fig:br1}\rm (l)}  &=& e\, 
 \sum_{B_x} 
D^{K^+B_j}_{B_x} D^{M_j p}_{B_x} S_4^{B_x}
\bm \sigma\cdot\bm \epsilon
O_4
\ ,\\
\label{eq:br-m}
V^j_{\ref{fig:br1}\rm (m)}  &=& - e\,
 \sum_{B_x} 
D^{K^+B_j}_{B_x} D^{M_j p}_{B_x} S_4^{B_x}
F_3
\left\{\left(1+{k'^0\over 2 M_B}\right)\bm \sigma\cdot\bm k' 
+{k'^0\over M_B} \bm \sigma\cdot \bm k_j\right\}
O_4
\ ,\\
\label{eq:brn-n}
V^j_{\ref{fig:br1}\rm (n)}  &=& e\, 
 \sum_{B_x} 
D^{K^+B_j}_{B_x} D^{M_j p}_{B_x} S_4^{B_x}
F_2
\left\{\left(1 - {k^0\over 2 M_B} \right) \bm \sigma\cdot\bm k 
+ {k^0\over M_B} \bm \sigma\cdot(\bm q+\bm k_j)\right\}
O_4
\nonumber \\
&& - e\, Q_{B_j} 
 \sum_{B_x} 
D^{K^+B_j}_{B_x} D^{M_j p}_{B_x} S_4^{B_x}
 {k^0\over 2 M_B} 
\bm \sigma\cdot\bm \epsilon
\bm \sigma\cdot\bm k_j
\ ,
\end{eqnarray}
where the summation $\sum_{B_x}$ runs over the octet baryons contained
in Eq.~(\ref{eq:baryon}).
We do not consider $\Lambda$-$\Sigma^0$ mixing mechanism due to the
anomalous magnetic moment in Eqs.~(\ref{eq:brn-k}) and (\ref{eq:brn-n})
because (i) it contributes to unimportant $K^+\Xi^-$ channel in Eq.~(\ref{eq:brn-k});
(ii) contribution of Eq.~(\ref{eq:brn-n}) is rather small.

\subsection{Vector meson exchange terms}

We introduce some building blocks as follows
to construct $V^j_x$ in this subsection:
\begin{eqnarray}
 O_5(a) &=&  
\left(a^0\bm q\cdot(\bm q -\bm a)-q^0\bm a\cdot (\bm q -\bm a)\right)\bm\sigma\cdot\bm \epsilon
+\left(q^0\bm\sigma\cdot\bm a-a^0\bm\sigma\cdot\bm q\right)(\bm q -\bm a)\cdot\bm \epsilon
\ , \\
O_6(a) &=& i \left(q^0 \bm\sigma\times\bm a - a^0 \bm\sigma\times\bm q \right)\cdot\bm\epsilon \ , \\
O_7 &=& 
\left\{\bm\sigma\cdot(\bm q - \bm k - \bm k_j) - {q^0-k^0-k_j^0\over 2 M_B} \bm\sigma\cdot(\bm k -
 \bm q - \bm k_j)
\right\}
\nonumber \\
&&\times
i \left(q^0\bm k\times\bm k_j + k^0\bm k_j\times\bm q + k_j^0\bm q\times\bm k
\right)\cdot\bm\epsilon \ , 
\end{eqnarray}
where $a$ is a four-vector. 
We also define the following propagators:
\begin{eqnarray}
S_5^{K^*} &=&  {1\over (q-k)^2 - M^2_{K^*} + i M_{K^*}\Gamma_{K^*}} \ , \\
S_6^{X} &=&  {1\over (q-k-k_j)^2 - M^2_{X} + i M_{X}\Gamma_{X}}  \ , \\
S_7^{V} &=&  {1\over (q-k_j)^2 - M^2_V + i M_{V}\Gamma_{V}}  \ , \\
S_8^{V} &=&  {1\over (k+k_j)^2 - M^2_V + i M_{V}\Gamma_{V}}  \ ,
\end{eqnarray}
where $X$ is a vector or pseudoscalar meson, and $V$ is a vector meson.
The mass of a particle $X$ is denoted by $M_X$, and its width by
$\Gamma_X$ for which we use the values from the PDG~\cite{pdg}.
For pseudoscalar particles, the width is set to zero.
We also define
\begin{eqnarray}
\bm J_B &=& 
{\bm p +\bm p'\over 2 M_B}
+i {(1+\kappa_{VpB_j})\bm\sigma\times(\bm p'-\bm p)\over 2M_B }\ , \\
J_1^0 &=& \bm k_j\times (\bm k - \bm q)\cdot \bm J_B\ ,\\
\bm J_1 &=& \bm k_j\times (\bm k - \bm q) + k_j^0 (\bm k - \bm q)\times\bm J_B
 +(q^0-k^0) \bm k_j\times \bm J_B \ ,\\
J_2^0 &=& \bm k\times (\bm k_j - \bm q)\cdot \bm J_B\ ,\\
\bm J_2 &=& \bm k\times (\bm k_j - \bm q) + k^0 (\bm k_j - \bm q)\times\bm J_B
 +(q^0-k_j^0) \bm k\times \bm J_B \ .
\end{eqnarray}
Thus, the photo-production amplitudes due to vector meson exchanges shown in Fig.~\ref{fig:vec} are
given by the following expressions:

\begin{eqnarray}
V^j_{\ref{fig:vec}\rm (a)}  &=& g_{\gamma K^{*-}K^+}\sum_{Y^0}g_{K^{*-}pY^0}D^j_{Y^0}S_1^{Y^0}S_5^{K^*}
O_1
\left\{\left(1+ {k'^0\over 2 M_B}\right) i \bm q\times\bm k\cdot\bm \epsilon
+{1+\kappa_{K^{*-}pY^0}\over 2 M_B}
O_5(k)
\right\}
\nonumber \\
&&
+  g_{\gamma K^{*-}K^+}\sum_{Y^0}g_{K^{*-}pY^0}D^j_{Y^0}S_5^{K^*} {k^0_j\over 2 M_B}
O_6(k)
\ , \\
V^j_{\ref{fig:vec}\rm (b)}  &=& i g_{\gamma K^{*-}K^+}g^j_{K^{*-}p}S_5^{K^*}
\left\{k^0 \bm\sigma\times\bm q - q^0 \bm\sigma\times\bm k
+{\bm\sigma\cdot(\bm k -\bm q- \bm k_j)\over 2 M_B} \bm q\times\bm k
\right\}\cdot\bm\epsilon\ , \\
V^j_{\ref{fig:vec}\rm (c)}  &=& g_{\gamma K^{*-}K^+}\sum_{B_x}g_{K^{*-}B_xB_j}D^{M_jB_x}_{p}S_4^{B_x}S_5^{K^*}
\nonumber \\
&&\times
\left[
\left\{
\left(1+ {q^0-k^0\over 2 M_B} \right) i \bm q\times\bm k
+ {q^0\over M_B} i \bm k_j\times\bm k
- {k^0\over M_B} i \bm k_j\times\bm q
\right\}\cdot\bm\epsilon
+{1+\kappa_{K^{*-}B_xB_j}\over 2 M_B}
O_5(k)
\right]
O_4
\nonumber \\
&&
+ g_{\gamma K^{*-}K^+}\sum_{B_x}g_{K^{*-}B_xB_j}D^{M_jB_x}_{p}S_5^{K^*}
{k^0_j\over 2 M_B}
O_6(k)
\ , \\
V^j_{\ref{fig:vec}\rm (d)}  &=& -2\, g_{\gamma K^{*-}K^+}\sum_{M_x}g_{K^{*-}M_xM_j}D^{M_xp}_{B_j}S_5^{K^*}S_6^{M_x}
O_7
\ , \\
V^j_{\ref{fig:vec}\rm (e)}  &=& \sum_{V,B_x}g_{\gamma VM_j}g_{VpB_x}D^{K^+B_j}_{B_x} S_3^{B_x}S_7^V
O_3
\nonumber \\
&&\times
\Bigg[
\left\{
\left(1+ {q^0+k_j^0\over 2 M_B}\right) i \bm q\times\bm k_j
+ {q^0\over M_B} i \bm k_j\times\bm k
+ {k_j^0\over M_B} i \bm k\times\bm q
\right\}\cdot\bm \epsilon
+{1+\kappa_{VpB_x}\over 2 M_B}
O_5(k_j)
\Bigg] 
\nonumber \\
&&
+ \sum_{V,B_x}g_{\gamma VM_j}g_{VpB_x}D^{K^+B_j}_{B_x} S_7^V
 {k^0\over 2 M_B}
O_6(k_j)
\ , \\
V^j_{\ref{fig:vec}\rm (f)}  &=& i \sum_{V}g_{\gamma VM_j}g^{K^+B_j}_{Vp}S_7^V
\left\{k_j^0 \bm\sigma\times\bm q - q^0 \bm\sigma\times\bm k_j
+{\bm\sigma\cdot(\bm k -\bm q- \bm k_j)\over 2 M_B} \bm q\times\bm k_j
\right\}\cdot\bm\epsilon\ , \\
V^j_{\ref{fig:vec}\rm (g)}  &=& 
\sum_{V,Y^0}g_{\gamma VM_j}g_{VY^0B_j}D^{K^+Y^0}_{p} S_2^{Y^0}S_7^V
\Bigg\{\left(1+ {q^0+k_j^0\over 2 M_B} \right) i \bm q\times\bm k_j
\cdot\bm\epsilon
+{1+\kappa_{VY^0B_j}\over 2 M_B}
O_5(k_j)
\Bigg\}
O_2
\nonumber \\
&&
+ \sum_{V,Y^0}g_{\gamma VM_j}g_{VY^0B_j}D^{K^+Y^0}_{p} S_7^V
{k^0\over 2 M_B}
O_6(k_j)
\ , \\
V^j_{\ref{fig:vec}\rm (h)}  &=& 2\, \sum_{V,M_x}g_{\gamma VM_j}g_{VM_xK^+}D^{M_xp}_{B_j}S_7^VS_6^{M_x}
O_7
\ , \\
V^j_{\ref{fig:vec}\rm (i)}  &=& -2\, \sum_{V,M_x}g_{\gamma VM_x}g_{VK^+M_j}D^{M_xp}_{B_j}S_8^VS_6^{M_x}
O_7
\ , \\
V^j_{\ref{fig:vec}\rm (j)}  &=& g_{\gamma K^{*-}K^+}\sum_{V_x}g_{K^{*-}V_xM_j}g_{V_xpB_j}
S_5^{K^*} S_6^{V_x}
\left(-q^0 \bm k\times \bm J_1 + k^0 \bm q\times \bm J_1 + J_1^0 \bm k\times \bm q
\right)\cdot \bm\epsilon
\ , \\
V^j_{\ref{fig:vec}\rm (k)}  &=& \sum_{V,V_x}g_{\gamma VM_j}g_{VV_xK^+}g_{V_xpB_j}
S_7^V S_6^{V_x}
\left(-q^0 \bm k_j\times \bm J_2 + k_j^0 \bm q\times \bm J_2 + J_2^0 \bm k_j\times \bm q
\right)\cdot \bm\epsilon
\ .
\end{eqnarray}

\subsection{Contact terms}
We present expressions for contact terms for $\gamma + p \to K^+ + M_j + B_j$.
\begin{eqnarray}
V^j_{c1}  &=& \lambda^j_1\, \left(\bm\sigma\cdot\bm k_j\bm\sigma\cdot\bm\epsilon
- \bm\sigma\cdot\bm k_j\bm\sigma\cdot\bm q 
{(2\bm k - \bm q)\cdot\bm \epsilon\over (k-q)^2-M_{K^+}^2}
\right)
\label{eq:c1}
\ , \\
V^j_{c2}  &=& \lambda^j_2\, \left(\bm\sigma\cdot\bm\epsilon \bm\sigma\cdot\bm k_j
- \bm\sigma\cdot\bm q  \bm\sigma\cdot\bm k_j
{(2\bm k - \bm q)\cdot\bm \epsilon\over (k-q)^2-M_{K^+}^2}
\right)
\label{eq:c2}
\ , \\
V^j_{c3}  &=& \lambda^j_3\, i (\bm\sigma\times\bm q)\cdot\bm\epsilon
\ , 
\label{eq:c3}
\end{eqnarray}
where $\lambda^j_n$ ($n=1,2,3$) are complex coupling constants that
depend on the total energy of the whole system.
Each term in Eqs.~(\ref{eq:c1})-(\ref{eq:c3}) is gauge invariant at
${\cal O}((1/M_B)^0)$.
\begin{figure}
\includegraphics[clip,width=0.8\textwidth]{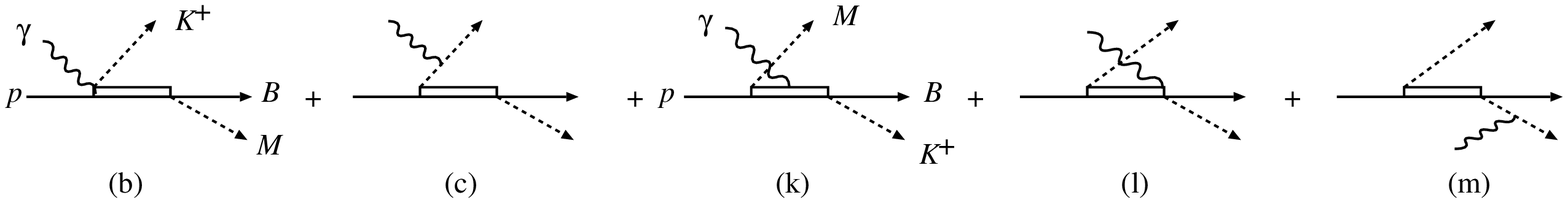}
\caption{\label{fig:c1} 
Nucleon (hyperon) resonance exchange diagrams. 
The diagrams (b), (c), (k), (l), and (m) are obtained from 
Figs.~\ref{fig:br1} (b), (c), (k), (l), and (m), respectively,
by replacing the intermediate nucleons (hyperons) with
nucleon (hyperon) resonances of $J^P=1/2^+$.
}
\end{figure}
A microscopic origin of the $V^j_{c1}$ term can be represented by diagrams shown in
Figs.~\ref{fig:c1} (b) and (c) where a hyperon resonance,
$\Lambda^*$ or $\Sigma^{0*}$, of $J^P=1/2^+$ ($J$:spin, $P$:parity) 
is exchanged.
The corresponding expression is given, retaining 
${\cal O}((1/M_B)^0)$ terms, as
\begin{eqnarray}
V^j_{\ref{fig:c1}\rm (b) + \ref{fig:c1}\rm (c)}  &=& 
\sum_{Y^*} C^{Y^*}_{\ref{fig:c1}\rm (b) + \ref{fig:c1}\rm (c)}
{\bm \sigma\cdot\bm k_j \bm \sigma\cdot\bm \epsilon 
- \bm \sigma\cdot\bm k_j \bm \sigma\cdot\bm k' 
{(2\bm k - \bm q)\cdot\bm \epsilon\over (k-q)^2-M_{K^+}^2}
 \over q+E_p(p)-E_{K^+}(k)- M_{Y^*} + i \Gamma_{Y^*}/2} \ ,
\end{eqnarray}
where $M_{Y^*}$ and $\Gamma_{Y^*}$ are the mass and width of the
exchanged hyperon resonance, respectively. 
The constant $C^{Y^*}_{\ref{fig:c1}\rm (b) + \ref{fig:c1}\rm (c)}$
is the product of coupling constants of $K^+pY^*$, $MBY^*$ and $e$ (electric charge).
We can derive Eq.~(\ref{eq:c1}) by putting the denominator and coupling
together into the $W$-dependent complex coupling $\lambda^j_1$.
Equation~(\ref{eq:c2}) can be derived in a similar way from the diagrams
of Figs.~\ref{fig:c1} (l) and (m), and 
Eq.~(\ref{eq:c3}) from Figs.~\ref{fig:c1} (k); for the latter, only the
gauge-invariant piece, a term proportional to 
$i (\bm\sigma\times\bm q)\cdot\bm\epsilon$, is retained.
Even though the contact terms of Eqs.~(\ref{eq:c1})-(\ref{eq:c3}) can be
related to the microscopic mechanisms shown in Fig.~\ref{fig:c1}, 
by fitting data, they also effectively simulate other mechanisms not
explicitly considered in our model.

\section{Fitting parameters}\label{app:fit-coup}

We present numerical values for parameters obtained from fitting the data.
For the model in which the chiral unitary amplitudes are implemented,
they are shown in 
TABLE~\ref{tab:asub}-\ref{tab:fit-param}.
Also, the cutoff value for the form factor of Eq.~(\ref{eq:ff}) 
obtained from the fit is $\Lambda=511$~MeV.

\begin{table}
\caption{\label{tab:asub} 
The subtraction constants, $a^j_{\alpha'}(\mu)$ ($\alpha'$=A,B,...,K), defined in
 Eq.~(\ref{eq:gpropdr}) for the chiral-unitary-based model.
The index $\alpha'$ has been introduced in TABLE~\ref{tab:sub}.
While the quoted values are from the chiral unitary model and are fixed,
the others are determined by fitting the data.
}
\renewcommand{\arraystretch}{1.2}
\tabcolsep=2.2mm
\begin{tabular}{c|ccccccccccc}\hline
$j$  & A& B& C& D& E& F& G& H& I& J& K \\\hline
$\bar{K} N$& ``$-$1.84''& $-$0.49&  1.38&  1.52& $-$0.83&  1.23& $-$2.12& $-$1.66&  0.39& $-$0.06&  1.58\\
$\pi\Sigma$& ``$-$2.00''&  1.44&  1.39& $-$1.69& $-$0.99&  0.56& $-$1.05& $-$1.55& $-$2.52& $-$0.80& $-$0.36\\
\hline
\end{tabular}
\end{table}

\begin{table}
\caption{\label{tab:fit-param} 
Coupling constants, $\lambda^j_n$ ($n=1,2,3$), defined in
 Eqs.~(\ref{eq:c1})-(\ref{eq:c3}) for the chiral-unitary-based model.
They are determined by fitting the data at each value of the total
 energy $W$.
}
\renewcommand{\arraystretch}{1.2}
\tabcolsep=2.2mm
\begin{tabular}{c|ccc|ccc}\hline
\hline
    & \multicolumn{3}{c|}{$W$=2.0~GeV}& \multicolumn{3}{c}{$W$=2.1~GeV}\\
$j$ & $\lambda^j_1$ & $\lambda^j_2$ & $\lambda^j_3$ & $\lambda^j_1$ & $\lambda^j_2$ & $\lambda^j_3$ 
\\\hline
 $K^- p$        &   18.6  +18.6$i$&   90.7  +49.4$i$&    1.1   $-$0.1$i$&    7.5  +17.5$i$&    2.2   +7.3$i$&    1.1   $-$0.0$i$\\
 $\bar{K}^0n$   &    6.2  +55.9$i$&   85.7 +177.9$i$&    0.2   +0.7$i$&    1.7   +6.2$i$&   $-$3.1   +5.9$i$&   $-$0.1   +0.3$i$\\
 $\pi^0\Sigma^0$&    3.4   +0.8$i$&    0.4   +0.0$i$&    0.0   +0.7$i$&   $-$0.2   $-$4.4$i$&    0.6   $-$1.0$i$&    0.1   +0.2$i$\\
 $\pi^+\Sigma^-$&    0.2   +1.5$i$&    0.4   $-$0.1$i$&    0.7   $-$0.0$i$&   $-$1.3   $-$0.7$i$&    0.1   +0.0$i$&    0.5   +0.4$i$\\
 $\pi^-\Sigma^+$&    9.3   +0.5$i$&    3.4   +0.1$i$&    0.1   +0.2$i$&    1.2   $-$4.7$i$&    2.3   +0.4$i$&    0.2   +0.2$i$\\
\hline
\end{tabular}

\vspace{2mm}
\begin{tabular}{c|ccc|ccc}\hline
    & \multicolumn{3}{c|}{$W$=2.2~GeV}& \multicolumn{3}{c}{$W$=2.3~GeV}\\
$j$ & $\lambda^j_1$ & $\lambda^j_2$ & $\lambda^j_3$ & $\lambda^j_1$ & $\lambda^j_2$ & $\lambda^j_3$ 
\\\hline
 $K^- p$        &   10.7  +12.1$i$&   $-$1.5   $-$5.6$i$&    0.5   +0.1$i$&    7.3   +4.4$i$&  $-$51.4  $-$52.4$i$&    0.5   +0.2$i$\\
 $\bar{K}^0n$   &    3.2  +12.8$i$&   19.4  +57.4$i$&    0.9   +0.2$i$&   $-$1.8   +4.3$i$&    1.2  +66.7$i$&    0.0   $-$0.1$i$\\
 $\pi^0\Sigma^0$&   $-$1.7   $-$1.9$i$&   $-$1.7   +0.5$i$&    0.3   +0.1$i$&   $-$0.5   $-$2.5$i$&    1.0   $-$0.4$i$&    0.2   $-$0.1$i$\\
 $\pi^+\Sigma^-$&   $-$1.1   +0.2$i$&   $-$0.0   +0.0$i$&    0.4   $-$0.3$i$&   $-$1.9   +0.7$i$&    0.0   $-$0.1$i$&    0.0   +0.4$i$\\
 $\pi^-\Sigma^+$&    4.3   $-$3.9$i$&    1.1   +0.5$i$&    0.1   +0.0$i$&    1.6   $-$3.5$i$&    1.3   +1.9$i$&    0.1   +0.2$i$\\
\hline
\end{tabular}
\end{table}

For the model in which the Brit-Wigner amplitudes (Eq.~(\ref{eq:bw})) are implemented,
the determined parameters are presented in
TABLE~\ref{tab:bw-param}-\ref{tab:fit-param-bw}.
Also, the cutoff value for the form factor of Eq.~(\ref{eq:ff}) is $\Lambda=560$~MeV.

\begin{table}[t]
\caption{\label{tab:bw-param} The Breit-Wigner parameters defined in Eq.~(\ref{eq:bw}).
}
\renewcommand{\arraystretch}{1.2}
\tabcolsep=2.2mm
\begin{tabular}{ccccc}\hline
$M_{BW}$ (MeV)& $\Gamma_{BW}$ (MeV)& $C^{BW}_{\bar{K}N,\bar{K}N}$&  $C^{BW}_{\bar{K}N,\pi\Sigma}$&  $C^{BW}_{\pi\Sigma,\pi\Sigma}$\\
\hline
 1412.0&   67.0& 7.54+3.00$i$& 1.61+0.49$i$& 0.55$-$1.66$i$\\
\hline
\end{tabular}
\end{table}

\begin{table}[t]
\caption{\label{tab:asub-bw} 
The subtraction constants, $a^j_{\alpha'}(\mu)$ ($\alpha'$=A,B,...,K.), defined in
 Eq.~(\ref{eq:gpropdr}) for the Breit-Wigner model.
}
\renewcommand{\arraystretch}{1.2}
\tabcolsep=2.2mm
\begin{tabular}{c|ccccccccccc}\hline
$j$  & A& B& C& D& E& F& G& H& I& J& K \\\hline
$\bar{K} N$& ``$-$1.84''& $-$0.72& $-$1.33&  2.59& $-$0.21&  1.79& $-$1.68& $-$1.66&  0.31& $-$1.82&  1.59\\
$\pi\Sigma$& ``$-$2.00''&  0.66& $-$2.47& $-$1.35&  3.00&  0.84& $-$0.81& $-$2.76& $-$2.85& $-$0.10& $-$1.53\\
\hline
\end{tabular}
\end{table}

\begin{table}[t]
\caption{\label{tab:fit-param-bw} 
Coupling constants, $\lambda^j_n$ ($n=1,2,3$), defined in
 Eqs.~(\ref{eq:c1})-(\ref{eq:c3}) for the Breit-Wigner model.
}
\renewcommand{\arraystretch}{1.2}
\tabcolsep=2.2mm
\begin{tabular}{c|ccc|ccc}\hline
\hline
    & \multicolumn{3}{c|}{$W$=2.0~GeV}& \multicolumn{3}{c}{$W$=2.1~GeV}\\
$j$ & $\lambda^j_1$ & $\lambda^j_2$ & $\lambda^j_3$ & $\lambda^j_1$ & $\lambda^j_2$ & $\lambda^j_3$ 
\\\hline
$K^- p$        &   84.1  +12.6$i$&  235.8  +56.9$i$&    1.4   $-$0.8$i$&   62.8   +2.6$i$&   67.2  +61.5$i$&    1.9   $-$1.7$i$\\
$\bar{K}^0n$   &    2.9   +6.6$i$&   13.5   +6.4$i$&    0.5   +0.5$i$&    0.7  +19.1$i$&   32.1  +16.4$i$&   $-$1.6   +1.2$i$\\
$\pi^0\Sigma^0$&    0.8   $-$0.1$i$&    0.5   $-$0.1$i$&    0.2   $-$0.6$i$&   $-$0.4   $-$2.5$i$&    1.8   $-$0.2$i$&    0.3   $-$0.3$i$\\
$\pi^+\Sigma^-$&   $-$1.6   +0.3$i$&    0.4   +0.0$i$&    0.3   $-$0.1$i$&   $-$1.4   $-$0.7$i$&    0.0   $-$0.2$i$&   $-$0.0   $-$0.2$i$\\
$\pi^-\Sigma^+$&    6.1   +0.1$i$&    3.2   +0.3$i$&   $-$0.2   $-$0.1$i$&   $-$0.1   $-$1.1$i$&    2.0   $-$0.4$i$&   $-$0.1   +0.5$i$\\
\hline
\end{tabular}

\vspace{2mm}
\begin{tabular}{c|ccc|ccc}\hline
    & \multicolumn{3}{c|}{$W$=2.2~GeV}& \multicolumn{3}{c}{$W$=2.3~GeV}\\
$j$ & $\lambda^j_1$ & $\lambda^j_2$ & $\lambda^j_3$ & $\lambda^j_1$ & $\lambda^j_2$ & $\lambda^j_3$ 
\\\hline
$K^- p$        &   27.5   +6.1$i$&   $-$1.1  +32.2$i$&    0.3   $-$0.7$i$&    0.9   +4.3$i$&   48.0  +71.6$i$&   $-$0.9   $-$1.0$i$\\
$\bar{K}^0n$   &   $-$7.1   $-$0.1$i$&   16.2   +6.9$i$&   $-$0.1   +0.3$i$&    6.2   +9.6$i$&   17.3   +6.9$i$&    0.7   +0.7$i$\\
$\pi^0\Sigma^0$&    0.1   $-$1.6$i$&    1.7   $-$0.3$i$&    0.2   $-$0.2$i$&    1.8   +0.5$i$&    0.6   +0.2$i$&    0.1   $-$0.0$i$\\
$\pi^+\Sigma^-$&   $-$0.8   +0.5$i$&    0.0   +0.0$i$&   $-$0.1   $-$0.2$i$&    0.8   +0.1$i$&    0.2   $-$0.2$i$&   $-$0.2   +0.2$i$\\
$\pi^-\Sigma^+$&    0.5   $-$0.6$i$&    1.2   +0.4$i$&    0.1   +0.4$i$&    2.4   +2.4$i$&   $-$1.2   $-$0.6$i$&   $-$0.1   +0.2$i$\\
\hline
\end{tabular}
\end{table}

A few notes are in order:
The subtraction
constants for channels other than $\bar K N$ and $\pi\Sigma$ are set to
the values used in chiral unitary amplitudes;
The subtraction constants in the chiral unitary
amplitudes are not adjusted in the fit;
the contact terms of Eqs.~(\ref{eq:c1})-(\ref{eq:c3}) couple to 
$\bar K N$ and $\pi\Sigma$ channels only.

\end{document}